\documentclass
[superscriptaddress,secnumarabic,amssymb,amsmath,nobibnotes,aps,prd,showkeys,showpacs,nofootinbib,onecolumn]{revtex4}%
\usepackage{graphicx}
\usepackage{epstopdf}
\usepackage{epsf}
\usepackage{bm}
\usepackage{amsmath}
\usepackage{amsfonts}
\usepackage{amssymb}%
\providecommand{\U}[1]{\protect\rule{.1in}{.1in}}

\newcommand{\be}{\begin{equation}}
\newcommand{\ee}{\end{equation}}

\newcommand{\mincir}{\raise
-3.truept\hbox{\rlap{\hbox{$\sim$}}\raise4.truept\hbox{$<$}\ }}
\newcommand{\magcir}{\raise
-3.truept\hbox{\rlap{\hbox{$\sim$}}\raise4.truept\hbox{$>$}\ }}

\begin{document}
\title{Extended analysis for the Evolution of the Cosmological history in
Einstein-Aether Scalar Field theory}
\author{Andronikos Paliathanasis}
\email{anpaliat@phys.uoa.gr}
\affiliation{Institute of Systems Science, Durban University of Technology, Durban 4000,
Republic of South Africa}

\begin{abstract}
We consider an Einstein-aether scalar field cosmological model where the
aether and the scalar field are interacting. The model of our consideration
consists the two different interacting models proposed in the literature by
Kanno et al. and by Donnelly et al. We perform an extended analysis for the
cosmological evolution as it is provided by the field equations by using
methods from dynamical systems; specifically, we determine the stationary
points and we perform the stability analysis of those exact solutions.

\end{abstract}
\date{\today}
\maketitle

\section{Introduction}

Gravitational theories where the Lorentz symmetry is violated have drawn the
attention of cosmologists the last years
\cite{lorv1,lorv2,lorv3,lorv4,carroll,ea1}. Ho\v{r}ava-Lifshitz theory is a
theory of quantum gravity which provides Einstein's GR as a limit.
Ho\v{r}ava-Lifshitz is a renormalization theory with consistent ultra-violet
behavior exhibiting an anisotropic Lifshitz scaling between time and space
\cite{hor3}.\ Ho\v{r}ava-Lifshitz theory has various applications in
gravitational theories from cosmological studies until compact stars
\cite{or1,or2,kir0,kir2,kir3,st1,st2,st3}.

There are various problems in Ho\v{r}ava-Lifshitz of major significant which
can not overpass the last years. For example, it has not been explained
detailed yet how the Lorentz invariance is restored on the low-energy problem,
indeed various proposals have been done on that problem based on the
coexistence of Ho\v{r}ava-Lifshitz with a Lorentz invariant matter sector with
controlled quantum corrections \cite{hl001,hl002}. In addition the complete
renormalization of Ho\v{r}ava-Lifshitz gravity have not been proved yet
\cite{rg01,rg04}. The renormalization of the projectable Ho\v{r}ava-Lifshitz
have been proved recently in \cite{rg02}, however while projectable
Ho\v{r}ava-Lifshitz theory has common physics properties with Einstein's GR,
the latter theory is not fully recovered by the projectable
Ho\v{r}ava-Lifshitz gravity \cite{Mukohyama:2009mz}. For an extended
discussion we refer the reader in \cite{st01a,st01b,st01c,wwhl} and references therein.

In the classical limit Ho\v{r}ava-Lifshitz is related with the Einstein-aether
gravitational theory. There is an one way equivalence, which means that every
solution of Einstein-aether theory is a also solution of Ho\v{r}ava-Lifshitz,
while the inverse it is not true \cite{jac01,jac01b}. The equivalency of the
two theories is not general true for other physical properties and results
which follow from the direct form of the field equations, such as the PPN
constraints \cite{st01c,st02}.

The kinematic quantities of a time-like vector field, known as aether field,
are introduced in the Einstein-Hilbert Action Integral, the selection of the
aether field defines the preferred frame. Important characteristics of the
Einstein-\ae ther theory are that it preserves locality and covariance; while
it contains Einstein's GR \cite{DJ,DJ2,Carru}.

Similarly with the Ho\v{r}ava-Lifshitz theory, Einstein-aether gravity has
many cosmological applications. Specifically it can describe various
cosmological phases such are the early-time and late-time acceleration phases
of the universe \cite{Barrow,in1,in2,in3,in4,in5,in6}. Other applications of
Einstein-aether theory in gravitational physics can be found in
\cite{in7,in8,in9,in10,in11,in12,in12a,in13,in14,in15,in16} and references therein.

In \cite{DJ}, Donnelly and Jacobson introduced a scalar field in the
Einstein-aether gravity such that the scalar field and the aether field to be
coupled and interact. In the model of Donnelly and Jacobson the interaction
term between the scalar field and the aether field is introduced by the
potential term. On the other hand, Kanno and Soda in \cite{soda1} considered a
scalar-aether interaction theory in which the interaction is introduced in the
coefficient terms of the aether field.

There are various studies in the literature of Einstein-aether gravity with a
scalar field. Static spherical symmetric solutions were studied in
\cite{in11,in12}. Anisotropic cosmological Einstein-aether scalar field models
studied in \cite{ks01,ks02,ks03}. Inflationary solutions for this theory
presented for the first time in \cite{Barrow}, while analysis of the evolution
of the dynamics for Einstein-aether scalar field theory presented in
\cite{ra1,anaether}. The analysis presented in \cite{ra1,anaether} based on
the Einstein-aether model proposed by Donnelly and Jacobson \cite{DJ}. In
\cite{ra1} the authors performed a complete analysis for the given
scalar-field interaction potential which was found in \cite{Barrow} and
provide inflationary solutions. The scalar-field interaction potential of
\cite{Barrow} is a power series in terms of exponential functions for the
scalar field and the expansion rate of the underlying
Friedmann--Lema\^{\i}tre--Robertson--Walker (FLRW) spacetime.

In this work we extend the analysis of \cite{anaether}, by considered a more
generic form of the scalar-field interaction model for the Einstein-aether
cosmology. Because of the form of the interaction which we assume our analysis
is valid and for the two different Einstein-aether scalar field theories
presented by Donnelly et al. \cite{DJ} and Kanno et al. \cite{soda1}. The
scope of this analysis is to understand the change of dynamics and the effects
in the cosmological history by the new interaction terms, as also, to compare
the two different Einstein-aether scalar field cosmological models in the case
that they can be comparable. The dynamics of the field equations and the
evolution of the cosmological history are studied by determine the
stationary/critical points of the field equations and determine their
stability. Such analysis has widely applied in the literature in various
cosmological models \cite{cop1,cop2,mo1,mo2,mo3,mo4,mo5,mo6,mo7,mo8,mo9,mo10}.
The plan of the paper is as follows.

In Section \ref{field} we briefly discuss the Einstein-aether scalar field
gravitational model and we present the cosmological field equations for the
model of our study. In Section \ref{sec3} we write the field equations by
using dimensionless variables by using the $H$-normalization. In addition we
define the four different possible families of stationary points. The main
results of this work are presented in Section \ref{sec4} where we derive the
stationary points for the four possible families of points, while we determine
the stability conditions. Finally, in\ Section \ref{sec5} we discuss our
results by comparing them with that of the analysis in~\cite{anaether} and we
draw our conclusions.

\section{Einstein-aether cosmology}

\label{field}

Einstein-\ae ther theory is a Lorentz violated gravitational theory which
consists GR coupled at second derivative order to a dynamical timelike unitary
vector field, the aether field, $u^{\mu}$. The vector field $u^{\mu}$ can be
thought as the four-velocity of the preferred frame.

The Action Integral of the Einstein-\ae ther theory is defined as \cite{Carru}%
\begin{equation}
S_{AE}=\int d^{4}x\sqrt{-g}R+\int d^{4}x\sqrt{-g}\left(  K^{\alpha\beta\mu\nu
}u_{\mu;\alpha}u_{\nu;\beta}+\lambda\left(  u^{c}u_{c}+1\right)  \right)
+S_{m}. \label{ae.01}%
\end{equation}
The first rhs term of the latter Action Integral is the Einstein-Hilbert
Lagrangian where $R$ is the Ricciscalar of the underlying geometric space with
metric $g^{\mu\nu};$ the second rhs term of (\ref{ae.01}) is introduced by the
\ae ther theory, $u^{\mu}$ is the \ae ther field, $\lambda$ is a Lagrange
multiplier and the tensor $K^{\alpha\beta\mu\nu}$ is defined as
\begin{equation}
K^{\alpha\beta\mu\nu}\equiv c_{1}g^{\alpha\beta}g^{\mu\nu}+c_{2}g^{\alpha\mu
}g^{\beta\nu}+c_{3}g^{\alpha\nu}g^{\beta\mu}+c_{4}g^{\mu\nu}u^{\alpha}%
u^{\beta}. \label{ae.02}%
\end{equation}
Parameters $c_{1},~c_{2},~c_{3}$ and $c_{4}$ are dimensionless constants and
define the coupling between the \ae ther field with gravity. Finally, the
third rhs term of (\ref{ae.01}) describes the matter source.

An equivalent way to write the Action Integral (\ref{ae.01}) is by using the
kinematic quantities $\theta,~\sigma,~\omega$ and $\alpha$ for the \ae ther
field, $u^{\mu}$. Hence, Action (\ref{ae.01}) is written as \cite{jac01}
\begin{equation}
S_{EA}=\int\sqrt{-g}dx^{4}\left(  R+\frac{c_{\theta}}{3}\theta^{2}+c_{\sigma
}\sigma^{2}+c_{\omega}\omega^{2}+c_{\alpha}\alpha^{2}\right)  +S_{m}.
\label{ae.03}%
\end{equation}
where parameters $c_{\theta},~c_{\sigma},~c_{\omega},~c_{a}$ are functions of
$c_{1},~c_{2},~c_{3}$ and $c_{4}$. As far as the values of the free parameters
of the theory, i.e. $c_{1},c_{2},~c_{3}$ and $c_{4}$ are concerned, they have
constrained before in literature. Observational data from binary pulsar
systems applied in \cite{con01}, while recently the gravitational wave event
GW170817 applied \cite{con02} to test the Einstein-aether theory and
constraint the free parameters. In addition, in \cite{con03} cosmological
constraints have been applied to constraint the Einstein-aether theory.

In tis work, the Action Integral of the matter source $S_{m}$ we assume that
it describes a scalar field minimally coupled to gravity but coupled to the
aether field, that is \cite{DJ}
\begin{equation}
S_{m}=\int\left(  \frac{1}{2}g^{\mu\nu}\phi_{,\mu}\phi_{,\nu}-V\left(
\theta,\sigma,\omega,\alpha,\phi\right)  \right)  . \label{ae.04}%
\end{equation}
where the interaction between the aether field and the scalar field is
described in the potential $V\left(  \theta,\sigma,\omega,\alpha,\phi\right)
$.

According to the cosmological principle the universe is considered to be
homogeneous and isotropic which means that it is described by the FLRW
spacetime. In addition we consider the spatial curvature to be zero, from
where it follows that the line element which describes the universe in large
scales is%
\begin{equation}
ds^{2}=-dt^{2}+a^{2}\left(  t\right)  \left(  dr^{2}+r^{2}d\theta^{2}%
+r^{2}\sin^{2}\theta d\phi^{2}\right)  . \label{fr0a}%
\end{equation}

As far as the aether field is concerned, we do the selection $u^{\mu}%
=\delta_{t}^{\mu}$, where someone calculates $\sigma=0,~\omega=0$ and
$\alpha=0$. Consequently, the Action Integral (\ref{ae.04}) is simplified as
follows
\begin{equation}
S_{EA}=\int\sqrt{-g}dx^{4}\left(  R+\frac{1}{2}g^{\mu\nu}\phi_{,\mu}\phi
_{,\nu}-V\left(  \theta,\phi\right)  \right)  , \label{fr01a}%
\end{equation}
where the term $\frac{c_{\theta}}{3}\theta^{2}$ has been absorbed in the
potential function $V\left(  \theta,\phi\right)  $. In addition, we assume
that the scalar field inherits the symmetries of the spacetime, that is,
$\phi=\phi\left(  t\right)  $.

The gravitational field equations for the latter Action integral and the line
element (\ref{fr0a}) are \cite{DJ}%
\begin{equation}
\frac{1}{3}\theta^{2}=\frac{1}{2}\dot{\phi}^{2}+V-\theta V_{\theta},
\label{fr1}%
\end{equation}%
\begin{equation}
\frac{2}{3}\dot{\theta}=-\dot{\phi}^{2}-\dot{\theta}V_{\theta\theta}-\dot
{\phi}V_{\theta\phi}, \label{fr2}%
\end{equation}%
\begin{equation}
\ddot{\phi}+\theta\dot{\phi}+V_{\phi}=0. \label{KG}%
\end{equation}
Recall that we have assumed $k=\frac{8\pi G}{c^{2}}=c=1$; however either if we
do not do that assumption, because in the following section we work with
dimensionless variables the physical constants play no role in our analysis.

We observe that in the limit $V\left(  \theta,\phi\right)  =V\left(
\phi\right)  $ or $V\left(  \theta,\phi\right)  =V\left(  \phi\right)
+\kappa\theta^{2}$, the field equations of general relativity are recovered,
while in the second case constant $\kappa$ change the gravitational constant
$k$.

A singular universe $a\left(  t\right)  =a_{0}t^{B}$ is recovered when the
scalar field potential $V\left(  \theta,\phi\right)  $ is of the form
\cite{Barrow}
\begin{equation}
V\left(  \phi,\theta\right)  =V_{0}e^{-\lambda\theta}+%
{\displaystyle\sum\limits_{r=0}^{n}}
V_{r}\theta^{r}e^{\frac{r-2}{2}\lambda\phi}, \label{pot.01a}%
\end{equation}
in which $V_{0},~V_{r}$ and $\lambda$ are constants, specifically $V_{r}$ are
the coupling constants of the the scalar field with the aether field. For the
scalar field the exact solution is $\phi\left(  t\right)  =\ln t^{\frac
{2}{\lambda}}$ and for the expansion rate $\theta\left(  t\right)  =3Bt^{-1}$
where $B=B\left(  V_{0},V_{r},\lambda\right)  $. In \cite{ra1} the latter
model studied in details, where the general cosmological evolution studied by
determining the stationary points and their stability.

In \cite{anaether} the cosmological viability of equations (\ref{fr1}%
)-(\ref{KG}) were studied for the potential form $V\left(  \theta,\phi\right)
=U\left(  \phi\right)  +Y\left(  \phi\right)  \theta$ where $U\left(
\phi\right)  $ and $Y\left(  \phi\right)  $ were arbitrary. In such
consideration $Y\left(  \phi\right)  $ is the coupling function between the
scalar field and the aether field. For this generic potential form exact
solutions also determined, from where we found that except the scaling
solution $a\left(  t\right)  =a_{0}t^{p}$ and the de Sitter universe $a\left(
t\right)  =a_{0}e^{H_{0}t}$, we can construct other kind of solutions such is
the $\Lambda$CDM universe with $a\left(  t\right)  =a_{0}\sinh^{\frac{2}{3}%
}\left(  \sqrt{\frac{2}{3}}\Lambda t\right)  $.

In this work we extend the analysis of \cite{anaether} by assuming the
potential form to be%
\begin{equation}
V\left(  \theta,\phi\right)  =U\left(  \phi\right)  +Y\left(  \phi\right)
\theta+\frac{1}{3}\left(  W^{2}\left(  \phi\right)  -1\right)  \theta^{2}.
\label{pot.011}%
\end{equation}

By replacing potential (\ref{pot.011}) in (\ref{fr1})-(\ref{KG}) we find%
\begin{equation}
\frac{1}{3}W^{2}\left(  \phi\right)  \theta^{2}=\frac{1}{2}\dot{\phi}%
^{2}+U\left(  \phi\right)  , \label{l0.1}%
\end{equation}%
\begin{equation}
\frac{2}{3}W^{2}\left(  \phi\right)  \dot{\theta}=-\dot{\phi}^{2}-Y\left(
\phi\right)  \dot{\phi}+\frac{4}{3}WW_{\phi}\theta\dot{\phi}, \label{l0.2}%
\end{equation}%
\begin{equation}
\ddot{\phi}+\theta\dot{\phi}+U_{\phi}+Y_{\phi}\theta+\frac{1}{3}\left(
W^{2}\left(  \phi\right)  \right)  _{\phi}\theta^{2}=0. \label{l0.3}%
\end{equation}

The modified Friedmann equations, namely equations (\ref{l0.1}) and
(\ref{l0.2}), can be written in a equivalent tensor form%

\begin{equation}
W^{2}\left(  \phi\right)  G_{ab}=T_{ab}, \label{pot.05}%
\end{equation}
where $G_{ab}$ is the Einstein tensor, and $T_{ab}$ is the energy momentum
tensor which describes the effective fluid source written as%
\begin{equation}
T_{ab}=\rho_{\phi}u_{a}u_{b}+p_{\phi}h_{ab}, \label{pot.06}%
\end{equation}
in which $h_{ab}=g_{ab}+u_{a}u_{b}$ is the projective tensor and $\rho_{\phi
}~$and $p_{\phi}$ are the effective energy density and pressure components
defined as%
\begin{equation}
\rho_{\phi}=\frac{1}{2}\dot{\phi}^{2}+U\left(  \phi\right)  ~,~p_{\phi}%
=\frac{1}{2}\dot{\phi}^{2}-U\left(  \phi\right)  +\dot{\phi}Y_{,\phi}+\frac
{4}{3}WW_{,\phi}\theta\dot{\phi}. \label{pot.07}%
\end{equation}

At this point, it is important to mentioned that while we consider the
scalar-aether model proposed in \cite{DJ}, for the function form
(\ref{pot.011}) of the unknown potential, the field equations of our model for
$Y\left(  \phi\right)  =0,$ reduce to the model of Kanno and Soda
\cite{soda1}. Hence, from the following analysis we are able to compare the
dynamical evolution of the two different theories.

From (\ref{pot.05}),\ we see that the term provides the effects of a variable
gravitational \textquotedblleft constant\textquotedblright\ $k$, that is
$k_{eff}=\left(  W^{2}\left(  \phi\right)  \right)  ^{-1}$, a similar
behaviour with the Scalar-tensor theories. While the scalar field is minimally
coupled to gravity it is interacting with the aether field, in which the
latter is coupling with gravity.

However, while scalar-tensor theories admit a minisuperspace description that
it is not true for this specific model. The energy density of the effective
fluid is that of the minimally coupled scalar field, while the pressure
$p_{\phi}$ differs with the additional terms to follow by the coupling
components of the scalar field with the aether field.

Finally, because of the $k_{eff}=\left(  W^{2}\left(  \phi\right)  \right)
^{-1}~$term we expect a difference on the physical evolution of the system
with the previous studied model in \cite{anaether} where the potential was
considered to be $V\left(  \theta,\phi\right)  =U\left(  \phi\right)
+Y\left(  \phi\right)  \theta$.

\section{Dimensionless variables}

\label{sec3}

In order to study the general evolution of the field equations (\ref{l0.1}%
)-(\ref{l0.3}) we work with the dimensionless variables defined as
\cite{cop1,cop2}%
\begin{equation}
x=\sqrt{\frac{3}{2}\frac{\dot{\phi}^{2}}{W^{2}\theta^{2}}}~,~~y=\sqrt
{\frac{3U}{W^{2}\theta^{2}}}~,~\lambda=\frac{U_{,\phi}}{U}~,~\xi=\sqrt{2}%
\frac{Y_{,\phi}}{\sqrt{U}}~,~~\zeta=2\frac{W_{,\phi}}{W}.
\end{equation}

In the new variables, the field equations are written as the following
algebraic-differential system%
\begin{align}
\frac{dx}{d\tau}  &  =\frac{1}{6}\left(  x^{2}-1\right)  \left(  3x+2\sqrt
{6}\zeta\right)  -\frac{1}{6}y^{2}\left(  3x+\sqrt{6}\lambda\right)  +\frac
{1}{2}\left(  x^{2}-1\right)  y\xi,\label{ds.01}\\
\frac{dy}{d\tau}  &  =y^{2}\left(  \left(  1-y^{2}\right)  +\frac{1}%
{3}x\left(  3\left(  x+y\xi\right)  +\sqrt{6}\right)  \left(  \lambda+\sqrt
{6}\zeta\right)  \right)  ,\label{ds.02}\\
\frac{d\lambda}{d\tau}  &  =\sqrt{\frac{2}{3}}x\lambda\left(  \zeta
+\lambda\left(  \Gamma^{\left(  \lambda\right)  }\left(  \lambda\right)
-1\right)  \right)  ,\label{ds.03}\\
\frac{d\xi}{d\tau}  &  =\frac{\sqrt{3}}{6}x\xi\left(  2\xi\Gamma^{\left(
\xi\right)  }\left(  \xi\right)  -\sqrt{2}\lambda\right)  ,\label{ds.04}\\
\frac{d\zeta}{d\tau}  &  =\frac{\sqrt{6}}{3}x\Gamma^{\left(  \zeta\right)
}\left(  \zeta\right)  , \label{ds.05}%
\end{align}
with algebraic constraint equation%
\begin{equation}
1-x^{2}-y^{2}=0. \label{ds.07}%
\end{equation}

The new independent variable $\tau$ is defined as $\frac{d\tau}{dt}=\theta$,
that is $\tau=\frac{1}{3}\ln a$ and describes the number of e-folds while
functions $\Gamma^{\left(  \lambda\right)  }\left(  \lambda\right)
,~\Gamma^{\left(  \xi\right)  }\left(  \xi\right)  $ and $\Gamma^{\left(
\zeta\right)  }\left(  \zeta\right)  $ are defined as%
\begin{equation}
\Gamma^{\left(  \lambda\right)  }\left(  \lambda\right)  =\frac{U_{\phi\phi}%
U}{U_{\phi}^{2}}~,~\Gamma^{\left(  \xi\right)  }\left(  \xi\right)
=\frac{Y_{\phi\phi}\sqrt{U}}{Y_{\phi}^{2}}\text{ and }\Gamma^{\left(
\zeta\right)  }=\zeta^{2}\frac{W_{\phi\phi}W}{W_{\phi}^{2}}. \label{ds.08}%
\end{equation}

In the new coordinates, the equation of state parameter for the total fluid
$w_{tot}$ is written as%
\begin{equation}
w_{tot}=x^{2}-y^{2}+xy\xi+\frac{4\sqrt{6}}{3}x\zeta. \label{ds.09}%
\end{equation}
One can conclude that equations (\ref{ds.01})-(\ref{ds.05}) have more degrees
of freedom than the field equations in the original variables of $\left\{
\theta,\phi\right\}  $. However that it is not true since equations
(\ref{ds.01})-(\ref{ds.05}) are not independent. Specifically variables
$\lambda,~\xi,~\zeta$ are not independent and in general one can always write
locally $\phi=\phi\left(  \lambda\right)  ,~$such that $\xi=\xi\left(
\lambda\right)  $ and $\zeta=\zeta\left(  \lambda\right)  $. In that case, the
independent equations of the dynamical system are equations (\ref{ds.01}),
(\ref{ds.02}) and (\ref{ds.03}). In addition when $\zeta=0$, that is $W\left(
\phi\right)  =const$. $\ $we see that the latter dynamical system reduces to
the one of \cite{anaether} as expected.

Before we continue with the rest of our analysis we present the different
families of stationary points. When variables $\lambda,~\xi$ and $\zeta$ are
constants, that is, $U\left(  \phi\right)  =U_{0}e^{\lambda\phi},~Y\left(
\phi\right)  =Y_{0}-\frac{\sqrt{2}}{4}\xi e^{-\frac{\lambda}{2}\phi}$ and
$W\left(  \phi\right)  =W_{0}e^{\frac{\zeta}{2}\phi}$, then the rhs of
equations (\ref{ds.03}), (\ref{ds.04}) and (\ref{ds.05}) are identical zero,
and the dynamical system is reduced to the two equations (\ref{ds.01}) and
(\ref{ds.02}). The stationary points of that system we call that they belong
to Family A. The stationary points which form the Family B are those of the
dynamical system (\ref{ds.01}), (\ref{ds.02}) and (\ref{ds.04}) where
$\lambda=const$. such that $\phi=\phi\left(  \xi\right)  $ and $\zeta
=\zeta\left(  \xi\right)  $.

The third family of points, namely Family C, it is consisted by the stationary
points of the dynamical system (\ref{ds.01}), (\ref{ds.02}) and (\ref{ds.05})
in which $\lambda=const$. and $\xi=cons\not t  ,$ such that $\phi=\phi\left(
\zeta\right)  $. However, for $U\left(  \phi\right)  \neq U_{0}e^{\lambda\phi
}$, such that $\lambda$ is a varying function, and $\phi=\phi\left(
\lambda\right)  $, then we end with the dynamical system (\ref{ds.01}),
(\ref{ds.02}) and (\ref{ds.03}) whose stationary points form the Family D.

Therefore, we conclude that points of Family $A$ are defined on the\ the
two-dimensional space $A=\left(  A_{x},A_{y}\right)  $, while points of
Families B, C and D are defined in the three-dimensional spaces $B=\left(
B_{x},B_{y},B_{\xi}\right)  ,~C=\left(  C_{x},C_{y},C_{\zeta}\right)  $ and
$D=\left(  D_{x},D_{y},D_{\lambda}\right)  $ respectively. However, from the
constraint equation (\ref{ds.07}) all the points in the plane $x-y$ are on the
border on the unitary circle, which means that the each dynamical system can
be reduced by one-dimension.

\section{Cosmological evolution}

\label{sec4}

In this section we present the stationary points and their stability for the
dynamical systems that we defined above; while we discuss the physical
quantities of the exact solutions at the stationary points.

\subsection{Family A}

The two dimensional dynamical system (\ref{ds.01}), (\ref{ds.02}) admits the
following four stationary points $\left(  A_{x},A_{y}\right)  $ which satisfy
the constraint equation (\ref{ds.07}),%
\begin{equation}
A_{1}^{\pm}=\left(  \pm1,0\right)  ~,~A_{2}^{\pm}=\left(  \frac{-2\sqrt
{6}\left(  2\zeta+\lambda\right)  \pm\sqrt{3\xi^{4}-6\left(  \left(
2\zeta+\lambda\right)  ^{2}-6\right)  \xi^{2}}}{3\left(  4+\xi^{2}\right)
},1-\left(  A_{2\left(  x\right)  }^{\pm}\right)  ^{2}\right)  .
\end{equation}
We observe that there are two families of points, the $A_{1}^{\pm}$ and
$A_{2}^{\pm}$ which include mirror points in the unitary circle.

Points $A_{1}^{\pm}$ describe universes where only the kinetic part of the
scalar field contributes the energy density of the effective fluid. The total
equation of state parameter is calculated%
\begin{equation}
w_{tot}\left(  A_{1}^{\pm}\right)  =1\pm\frac{4\sqrt{6}}{3}\zeta,
\label{ea.02}%
\end{equation}
from where we infer that the coupling term $\theta^{2}W\left(  \phi\right)  $
contributes in the pressure term also such that to modify the equation of
state parameter from that of stiff fluid as in the case of General Relativity.
From (\ref{ea.02}) we observe that now $w_{tot}$ can take values lower that
$-1$. If we constraint $\left\vert w_{tot}\left(  A_{1}^{\pm}\right)
\right\vert \leq1$, then we find that $\zeta\left(  A_{1}^{+}\right)
\in\left[  -\frac{1}{2}\sqrt{\frac{3}{2}},0\right]  ,$ for point $A_{1}^{+}$
and $\zeta\left(  A_{1}^{-}\right)  \in\left[  0,\frac{1}{2}\sqrt{\frac{3}{2}%
}\right]  ,$ for point $A_{1}^{-}$.

\begin{figure}[ptb]
\centering \includegraphics[width=0.4\textwidth]{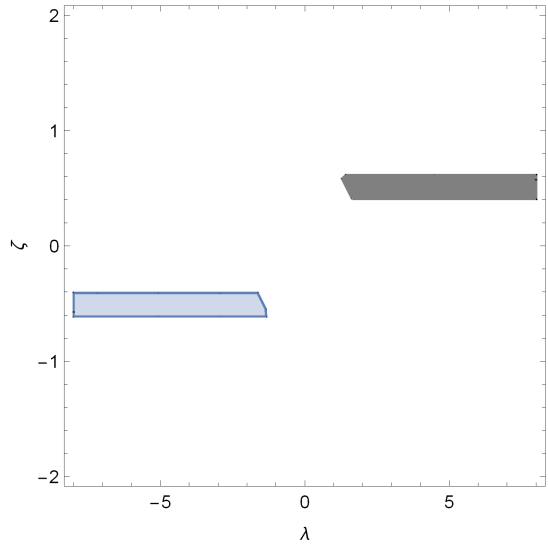} \newline%
\caption{Region plot in the space of variables $\left\{  \lambda
,\zeta\right\}  $ where the exact solutions at points $A_{1}^{\pm}$ describe
stable accelerated universes. The left region correspond to point $A_{1}^{+}$,
while the right region correspond to point $A_{1}^{-}$. }%
\label{fig1}%
\end{figure}

In order to study the stability of the stationary point we replace
$x=\cos\omega$ and \thinspace\thinspace$y=\sin\omega$ from where we find the
equation%
\begin{equation}
\frac{d\omega}{d\tau}=\frac{2\zeta+\lambda}{\sqrt{6}}\cos\omega+\cos\left(
2\omega\right)  +\frac{\xi}{2}\sin\left(  2\omega\right)  , \label{ea.03}%
\end{equation}
where points $A_{1}^{\pm}$ correspond to $\omega_{1}^{+}=2\pi N$ and
$\omega_{1}^{-}=\pi+2\pi N$, where $N$ is an integer number. Hence, the
linearized equation $\omega=\omega_{1}^{\pm}+\delta\omega$ around the
stationary points is
\begin{equation}
\frac{d\left(  \delta\omega\right)  }{d\tau}=\left(  1\pm\frac{2\zeta+\lambda
}{\sqrt{6}}\right)  \delta\omega,
\end{equation}
from where it follows that points $A_{1}^{\pm}$ are stable when $\left(
1\pm\frac{2\zeta+\lambda}{\sqrt{6}}\right)  <0$, that is $\zeta\left(
A_{1}^{+}\right)  <-\frac{6+\sqrt{6}\lambda}{2\sqrt{6}}$ for $A_{1}^{+}$ and
$\zeta\left(  A_{1}^{-}\right)  <\frac{6-\sqrt{6}\lambda}{2\sqrt{6}}$ for
$A_{1}^{-}.$ Now if we assume that the points to describe accelerated
solutions, that is, $w_{tot}\left(  A_{1}^{\pm}\right)  <-\frac{1}{3}$ and be
attractors we find for point $A_{1}^{+},\left\{  \lambda<-2\sqrt{\frac{2}{3}%
},~-\frac{1}{2}\sqrt{\frac{3}{2}}\leq\zeta\leq-\frac{\sqrt{6}}{6}\right\}
\cup\left\{  -2\sqrt{\frac{2}{3}}<\lambda<-\sqrt{\frac{3}{2}},~-\frac{1}%
{2}\sqrt{\frac{3}{2}}\leq\zeta<-\frac{6+\sqrt{6}\lambda}{2\sqrt{6}}\right\}
\cup\left\{  \lambda=-2\sqrt{\frac{2}{3}},~-\frac{1}{2}\sqrt{\frac{3}{2}}%
\leq\zeta<-\frac{\sqrt{6}}{6}\right\}  $, while for point $A_{1}^{-}$ we
find~$\left\{  \sqrt{\frac{3}{2}}\leq\lambda\leq2\sqrt{\frac{2}{3}}%
,\frac{6-\sqrt{6}\lambda}{2\sqrt{6}}<\zeta\leq\frac{1}{2}\sqrt{\frac{3}{2}%
}~\right\}  \cup\left\{  \lambda>2\sqrt{\frac{2}{3}},\frac{\sqrt{6}}{6}%
<\zeta\leq\frac{1}{2}\sqrt{\frac{3}{2}}\right\}  .$ The latter regions are
plotted in Fig. \ref{fig1}.

Points $A_{2}^{\pm}$ depend on the three constants of the problem. Points are
physical accepted when $\xi^{2}\left(  \xi^{2}-2\left(  \left(  2\zeta
+\lambda\right)  ^{2}-6\right)  \right)  \geq0$, that is when $\xi^{2}%
\geq2\left(  \left(  2\zeta+\lambda^{2}\right)  -6\right)  $, or when $\xi=0$.
The equation of state parameter at the points is calculated to be
\begin{equation}
w_{tot}\left(  A_{2}^{\pm}\right)  =-1-\left(  2\zeta-\lambda\right)
\frac{4\left(  2\zeta+\lambda\right)  \pm\sqrt{2}\sqrt{6\xi^{4}-4\left(
\left(  2\zeta+\lambda\right)  ^{2}-6\right)  }}{3\left(  4+\xi^{2}\right)  }.
\end{equation}
In order to conclude about the stability of the stationary points we reduce
the dynamical system to one equation with dependent variable the
$\omega\left(  \tau\right)  $. Hence, the linearized system around the
stationary points $\omega_{2}^{\pm}$, are%
\begin{equation}
\frac{d\left(  \delta\omega\right)  }{d\tau}=\frac{\sqrt{3\left(  4+\xi
^{2}\right)  -2\left(  2\zeta+\lambda\right)  ^{2}}\left(  \sqrt{2}\left(
2\zeta+\lambda\right)  \mp2\sqrt{3\left(  4+\xi^{2}\right)  -2\left(
2\zeta+\lambda\right)  ^{2}}\right)  }{6\left(  4+\xi^{2}\right)  }%
\delta\omega,
\end{equation}
from where it follows that the point $A_{2}^{+}$ is stable when $\left\{
\xi<0,~-\frac{\sqrt{6\left(  4+\xi^{2}\right)  }}{2}<Z<-\sqrt{6}\right\}
\cup\left\{  \xi>0,\sqrt{6}<Z<\sqrt{\frac{3}{2}\left(  4+\xi^{2}\right)
}\right\}  $ in which $Z=2\zeta+\lambda$. On the other hand, point $A_{2}^{-}$
is an attractor when $\left\{  \xi>0,~-\sqrt{6}<Z<\frac{\sqrt{6\left(
4+\xi^{2}\right)  }}{2}\right\}  \cup\left\{  \xi\leq0,-\sqrt{\frac{3}%
{2}\left(  4+\xi^{2}\right)  }<Z<\sqrt{6}\right\}  $.

\begin{figure}[ptb]
\centering\includegraphics[width=0.4\textwidth]{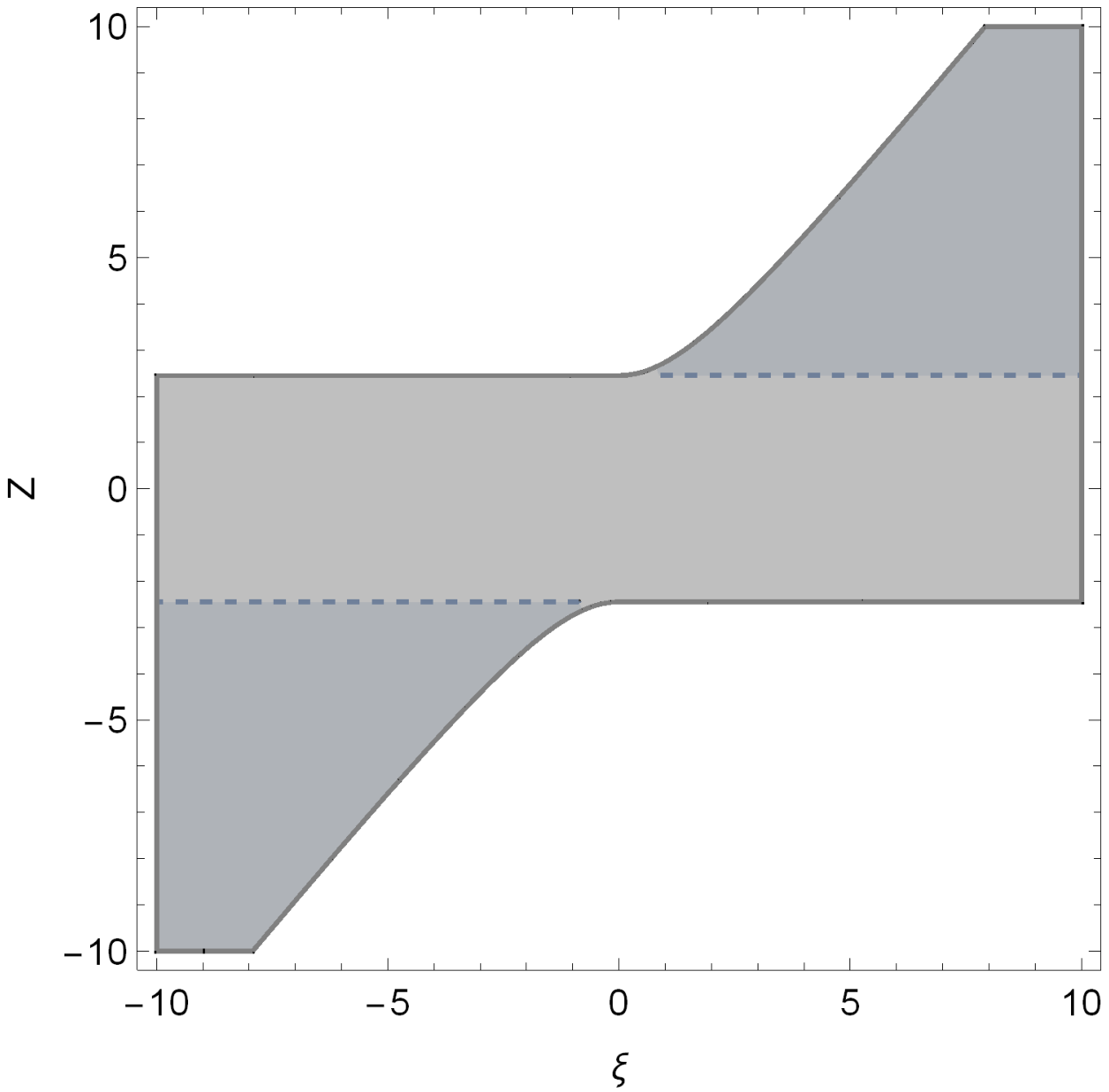} \newline%
\caption{Region plot in the space of variables $\left\{  \xi,\left(
2\zeta+\lambda\right)  \right\}  $ where the exact solutions at points
$A_{2}^{\pm}$ are stable. Blue area corredpond to the values where point
$A_{2}^{+}$ is stable, while gray area is for point $A_{2}^{-}$. }%
\label{fig2}%
\end{figure}

In Fig. \ref{fig2}, we present the region in the three-dimensional space of
the free parameters $\left\{  \lambda,\xi,\zeta\right\}  $ in which the points
$A_{2}^{\pm}$ are attractors, and when the solution at the point is stable and
describes an accelerated universe.

\subsection{Family B}

From the rhs of equations (\ref{ds.01}), (\ref{ds.02}), (\ref{ds.04}) we find
the stationary points~$B=\left(  B_{x},B_{y},B_{\xi}\right)  $ which belong to
family B, they are
\begin{align}
B_{1}^{\pm}  &  =\left(  \pm1,0,0\right)  ~,\\
B_{2}^{\pm}  &  =\left(  \pm1,0,\xi_{0}\right)  ~,~\sqrt{2}\Gamma^{\left(
\xi\right)  }\left(  \xi_{0}\right)  \xi_{0}=\lambda~,~\\
B_{3}^{\pm}  &  =\left(  -\frac{2\sqrt{2}\left(  2\left(  \xi_{0}\right)
+\lambda\right)  \pm\sqrt{3\xi_{0}^{4}-2\left(  \left(  2\zeta\left(  \xi
_{0}\right)  +\lambda\right)  ^{2}-6\right)  \xi_{0}^{2}}}{\sqrt{3}\left(
4+\xi_{0}^{2}\right)  },1-\left(  B_{4x}^{\pm}\right)  ^{2},\xi_{0}\right)
~,~\sqrt{2}\Gamma^{\left(  \xi\right)  }\left(  \xi_{0}\right)  \xi
_{0}=\lambda~,\\
B_{4}^{\pm}  &  =\left(  0,1,\pm\sqrt{\frac{2}{3}}\left(  2\zeta\left(
\xi_{0}\right)  +\lambda\right)  \right)  ~,\\
B_{5}^{\pm}  &  =\left(  -\frac{2\zeta+\lambda}{\sqrt{6}},\pm\frac
{\sqrt{6-\left(  2\zeta\left(  \xi_{0}\right)  +\lambda^{2}\right)  }}%
{\sqrt{6}},0\right)  .
\end{align}

Points $B_{1}^{\pm}$, $B_{2}^{\pm}$ describe the same physical physical
solution as points $A_{1}^{\pm}$ where the equation of state for the effective
fluid is $w_{tot}\left(  B_{1}^{\pm},B_{2}^{\pm}\right)  =1\pm\frac{4\sqrt{6}%
}{3}\zeta$.

At the points $B_{1}^{\pm}$ there is not any contribution in the evolution of
the field equation by the term of $Y\left(  \phi\right)  \theta$ since
$\xi\left(  B_{2}^{\pm}\right)  =0$. That is not true for the points
$B_{2}^{\pm}$ where in general $\xi\left(  B_{2}^{\pm}\right)  \neq0$ but
because $y\left(  B_{2}^{\pm}\right)  =0$ the contribution of the $Y\left(
\phi\right)  \theta$ is neglected. In addition it is important to note that
points $B_{2}^{\pm}$ exist if and only if there exist a real solution in the
algebraic equation $\sqrt{2}\Gamma^{\left(  \xi\right)  }\left(  \xi
_{0}\right)  \xi_{0}=\lambda~$.

In addition points $B_{3}^{\pm}$ describe the same physical solution with that
of points $A_{2}^{\pm}$ respectively, while $w_{tot}\left(  B_{3}^{\pm
}\right)  =w_{tot}\left(  A_{2}^{\pm}\right)  $.

The two new sets of points, namely $B_{4}^{\pm}$ and $B_{5}^{\pm}$ are of
special interest since provide addition phases in the cosmological evolution.
Points $B_{4}^{\pm}$ describe de Sitter solutions since $w_{tot}\left(
B_{3}^{\pm}\right)  =-1$. That is, the effective fluid source the stationary
points it mimics the cosmological constant. On the other hand, the stationary
points $B_{5}^{\pm}$ provide scaling solutions which can be seen as
generalized solutions of that of the scaling solution for the minimally
coupled scalar field in General Relativity. Indeed the limit of General
Relativity is recovered at the limit where $\zeta\rightarrow0$.

\subsubsection{Stability analysis}

We proceed by studying the stability of the stationary points. To do that
prefer to reduce the dynamical by one dimension by applying the change of
variables $x=\cos\omega,~y=\sin\omega$, where system (\ref{ds.01}),
(\ref{ds.02}), (\ref{ds.04}) is reduced to the following set of equations
\begin{align}
\frac{d\omega}{d\tau}  &  =\frac{2\zeta+\lambda}{\sqrt{6}}\sin\omega+\frac
{1}{2}\sin\left(  2\omega\right)  +\frac{\xi}{2}\sin^{2}\omega,\\
\frac{d\xi}{d\tau}  &  =\frac{\sqrt{3}}{6}\xi\cos\omega\left(  2\xi
\Gamma^{\left(  \xi\right)  }\left(  \xi\right)  -\sqrt{2}\lambda\right)  .
\end{align}

For points $B_{1}^{\pm}$ the eigenvalues of the linearized system are found to
be%
\begin{equation}
e_{1}\left(  B_{1}^{\pm}\right)  =\mp\frac{\lambda}{\sqrt{6}}~,~e_{2}\left(
B_{1}^{\pm}\right)  =1\pm\frac{\sqrt{6}}{6}\left(  2\zeta+\lambda\right)  ,
\end{equation}
from where we can infer that $B_{1}^{+}$ is an attractor when $\lambda>0$ and
$\left(  2\zeta+\lambda\right)  <-\sqrt{6}$, while $B_{1}^{-}$ is an attractor
when $\lambda<0$ and $\left(  2\zeta+\lambda\right)  >\sqrt{6}$. \ 

For the stationary points $B_{2}^{\pm}$ the eigenvalues are found to be%
\begin{equation}
e_{1}\left(  B_{2}^{\pm}\right)  =1\pm\frac{\sqrt{6}}{6}\left(  2\zeta
+\lambda\right)  ~,~e_{2}\left(  B_{2}^{\pm}\right)  =\pm\frac{1}{\sqrt{3}%
}\left(  \frac{\sqrt{2}}{2}\lambda+\xi_{0}\Gamma_{,\xi}^{\left(  \xi\right)
}\left(  \xi_{0}\right)  \right)  .
\end{equation}
Hence, at the point $B_{2}^{+}$ the solution is stable, when $\left(
2\zeta+\lambda\right)  <-\sqrt{6}$ and $\frac{\sqrt{2}}{2}\lambda+\xi
_{0}\Gamma_{,\xi}^{\left(  \xi\right)  }\left(  \xi_{0}\right)  <0$. Recall
that $\sqrt{2}\Gamma^{\left(  \xi\right)  }\left(  \xi_{0}\right)  \xi
_{0}=\lambda$. In addition, point $B_{2}^{-}$ is a stable point when $\left(
2\zeta+\lambda\right)  >\sqrt{6}$ and $\frac{\sqrt{2}}{2}\lambda+\xi_{0}%
\Gamma_{,\xi}^{\left(  \xi\right)  }\left(  \xi_{0}\right)  >0$.

The eigenvalues of the linearized system at point $B_{3}^{\pm}$ are derived to
be%
\begin{align}
e_{1}\left(  B_{3}^{\pm}\right)   &  =-\frac{2\sqrt{2}\left(  2\zeta
+\lambda\right)  \mp\sqrt{3\xi_{0}^{4}-2\left(  \left(  2\xi_{0}%
+\lambda\right)  ^{2}-6\right)  \xi_{0}^{2}}}{6\left(  4+\xi_{0}^{2}\right)
}\left(  \sqrt{2}\lambda+2\xi_{0}^{2}\Gamma_{,\xi}^{\left(  \xi\right)
}\left(  \xi_{0}\right)  \right)  ,\\
e_{2}\left(  B_{3}^{\pm}\right)   &  =-1+\frac{\left(  2\zeta+\lambda\right)
\left(  4\left(  2\zeta+\lambda\right)  \mp\sqrt{2}\xi\sqrt{3\xi_{0}%
^{4}-2\left(  \left(  2\xi_{0}+\lambda\right)  ^{2}-6\right)  \xi_{0}^{2}%
}\right)  }{6\left(  4+\xi_{0}^{2}\right)  },
\end{align}
however in order to infer about the stability, parameter $\Gamma_{,\xi
}^{\left(  \xi\right)  }\left(  \xi_{0}\right)  ~$should be determined. Indeed
for $\sqrt{2}\lambda+2\xi_{0}^{2}\Gamma_{,\xi}^{\left(  \xi\right)  }\left(
\xi_{0}\right)  >0$ the solution at point $B_{3}^{+}$ is stable in the
following regions~when $\xi_{0}=0:\left\{  -\sqrt{6}<2\zeta+\lambda
<0~,~\lambda>0\right\}  $~while when $\xi_{0}\neq0:\left\{  -\sqrt{6}%
<2\zeta+\lambda<0\right\}  $ or $\left\{  \left(  2\zeta+\lambda\right)
>0,\sqrt{6}\left(  2\zeta+\lambda\right)  <3\xi_{0}~\right\}  ~$or$~\left\{
\left(  2\zeta+\lambda\right)  >0,\sqrt{6}\left(  2\zeta+\lambda\right)
<-3\xi_{0}~\right\}  $. On the other hand, when $\left(  \sqrt{2}\lambda
+2\xi_{0}^{2}\Gamma_{,\xi}^{\left(  \xi\right)  }\left(  \xi_{0}\right)
\right)  <0$ point $B_{3}^{+}$ is stable when $\xi_{0}=0:\left\{
0<2\zeta+\lambda<\sqrt{6},~\lambda<0\right\}  $ or $\xi_{0}\neq0:\left\{
\left(  2\zeta+\lambda\right)  <\sqrt{6}~,~\sqrt{6}\left(  2\zeta
+\lambda\right)  >3\xi_{0}~,~\xi_{0}<0\right\}  ~$or ~$\left\{  \left(
2\zeta+\lambda\right)  <\sqrt{6}~,~\sqrt{6}\left(  2\zeta+\lambda\right)
<-3\xi_{0}~,~\xi_{0}>0\right\}  \,\ $or $\left\{  2\zeta+\lambda>\sqrt
{6},~\sqrt{6}\left(  2\zeta+\lambda\right)  >3\xi_{0},~\sqrt{6}\left(
\sqrt{\left(  2\zeta+\lambda\right)  ^{2}-6}<3\xi_{0}\right)  \right\}  $ or
in the region $\left\{  2\zeta+\lambda>\sqrt{6},~\sqrt{6}\left(
2\zeta+\lambda\right)  >-3\xi_{0},~\sqrt{6}\left(  \sqrt{\left(
2\zeta+\lambda\right)  ^{2}-6}<-3\xi_{0}\right)  \right\}  $.

Similarly, when $\sqrt{2}\lambda+2\xi_{0}^{2}\Gamma_{,\xi}^{\left(
\xi\right)  }\left(  \xi_{0}\right)  >0$ point $B_{3}^{-}$ is an attractor
when$~\xi_{0}=0:\left\{  0<2\zeta+\lambda<\sqrt{6},~\lambda>0\right\}  $ or
$\xi_{0}\neq0:\left\{  2\zeta+\lambda<\sqrt{6},~\lambda<0\right\}  $ or
$\left\{  ~2\zeta+\lambda>0\right\}  \,$\ or $\left\{  \lambda<0,~2\zeta
+\lambda<0\right\}  \,\ $or $\left\{  \sqrt{6}\left(  2\zeta+\lambda\right)
<3\xi_{0}~,~\sqrt{6}\left(  2\zeta+\lambda\right)  <-3\xi_{0}\right\}  $.~In
addition when $\sqrt{2}\lambda+2\xi_{0}^{2}\Gamma_{,\xi}^{\left(  \xi\right)
}\left(  \xi_{0}\right)  <0$ point $B_{3}^{-}$ is stable when $\xi
_{0}=0:\left\{  2\zeta+\lambda>-\sqrt{6}\right\}  $ or $\xi_{0}\neq0:\left\{
\sqrt{6}\left(  2\zeta+\lambda\right)  >-3\xi,\sqrt{6}\left\vert \left(
2\zeta+\lambda\right)  \right\vert >3\left\vert \xi\right\vert \right\}  $ or
$\left\{  \sqrt{6}\left\vert 2\zeta+\lambda\right\vert >-3\xi,~\sqrt{6}%
\sqrt{\left(  2\zeta+\lambda\right)  ^{2}-6}<-3\xi\right\}  $ or $\left\{
\sqrt{6}\left\vert 2\zeta+\lambda\right\vert >3\xi,~\sqrt{6}\sqrt{\left(
2\zeta+\lambda\right)  ^{2}-6}<3\xi\right\}  $.

For the stationary points $B_{4}^{\pm}$ the eigenvalues are derived
\begin{align}
e_{1}\left(  B_{4}^{\pm}\right)   &  =-\frac{\left(  3+\sqrt{9\mp2\Gamma
_{2}\left(  \xi_{0}\right)  \left(  2\zeta\left(  \xi_{0}\right)
+\lambda\right)  \left(  3\sqrt{2}+4\sqrt{3}\zeta_{,\xi}\left(  \xi
_{0}\right)  \right)  +2\lambda\left(  2\zeta\left(  \xi_{0}\right)
+\lambda\right)  \left(  3+2\sqrt{6}\right)  \zeta_{,\xi}\left(  \zeta
_{0}\right)  }\right)  }{6},\\
e_{2}\left(  B_{4}^{\pm}\right)   &  =-\frac{\left(  3-\sqrt{9\mp2\Gamma
_{2}\left(  \xi_{0}\right)  \left(  2\zeta\left(  \xi_{0}\right)
+\lambda\right)  \left(  3\sqrt{2}+4\sqrt{3}\zeta_{,\xi}\left(  \xi
_{0}\right)  \right)  +2\lambda\left(  2\zeta\left(  \xi_{0}\right)
+\lambda\right)  \left(  3+2\sqrt{6}\right)  \zeta_{,\xi}\left(  \zeta
_{0}\right)  }\right)  }{6}.
\end{align}
From the latter eigenvalues and for $\zeta\left(  \xi_{0}\right)  =const$,
i.e. $\zeta\left(  \xi_{0}\right)  =\zeta_{0}$ we find that points $B_{4}%
^{\pm}$ are spiral attractors when $9+\left(  2\zeta_{0}+\lambda\right)
\left(  6\lambda-4\sqrt{3}\left(  2\zeta+\lambda\right)  \Gamma_{,\xi
}^{\left(  \xi\right)  }\left(  \xi_{0}\right)  \right)  \leq0$; while point
$B_{4}^{-}$ is also stable when $\left\{  \lambda=0,\zeta_{0}\neq0\text{ and
}\Gamma_{,\xi}^{\left(  \xi\right)  }\left(  \xi_{0}\right)  <0\right\}  $ or
$\left(  2\zeta+\lambda\right)  \Gamma_{,\xi}^{\left(  \xi\right)  }\left(
\xi_{0}\right)  +\sqrt{3}\lambda<0:\left\{  \lambda<0,~2\zeta_{0}%
+\lambda<0\right\}  $ or the region $\left\{  \lambda<0,~0<2\zeta_{0}%
+\lambda,8\zeta_{0}+\frac{3}{\lambda}+4\lambda\leq0\right\}  $ or $\left\{
\lambda>0,~8\zeta_{0}+\frac{3}{\lambda}+4\lambda>0,~2\zeta_{0}+\lambda
<0\right\}  $ or $\left\{  \lambda>0,2\zeta_{0}+\lambda>0\right\}  $ or
$\left(  2\zeta_{0}+\lambda\right)  \Gamma_{,\xi}^{\left(  \xi\right)
}\left(  \xi_{0}\right)  +\sqrt{3}\lambda>0:\left\{  3+4\lambda\left(
3\zeta_{0}+\lambda\right)  <0\right\}  $.

For the set of points $B_{5}^{\pm}$ we find the eigenvalues%
\begin{equation}
e_{1}\left(  B_{5}^{\pm}\right)  =\frac{\lambda}{6}\left(  2\zeta
+\lambda\right)  ,~e_{2}\left(  B_{5}^{\pm}\right)  \frac{1}{6}\left(  \left(
2\zeta+\lambda\right)  ^{2}-6\right)  ,
\end{equation}
from where we conclude that points $B_{5}^{\pm}$ are attractors for $\left\{
\zeta<-\sqrt{15},-2\sqrt{15}<2\zeta+\lambda<0\right\}  $ or $\left\{
-\sqrt{15}<\zeta<0,2\zeta<2\zeta+\lambda<0\right\}  $ or $\left\{
0<\zeta<\sqrt{15},~0<2\zeta+\lambda<2\zeta\right\}  $ or $\left\{  \zeta
>\sqrt{15},~0<2\zeta+\lambda<2\sqrt{15}\right\}  .$

\begin{figure}[ptb]
\centering\includegraphics[width=0.8\textwidth]{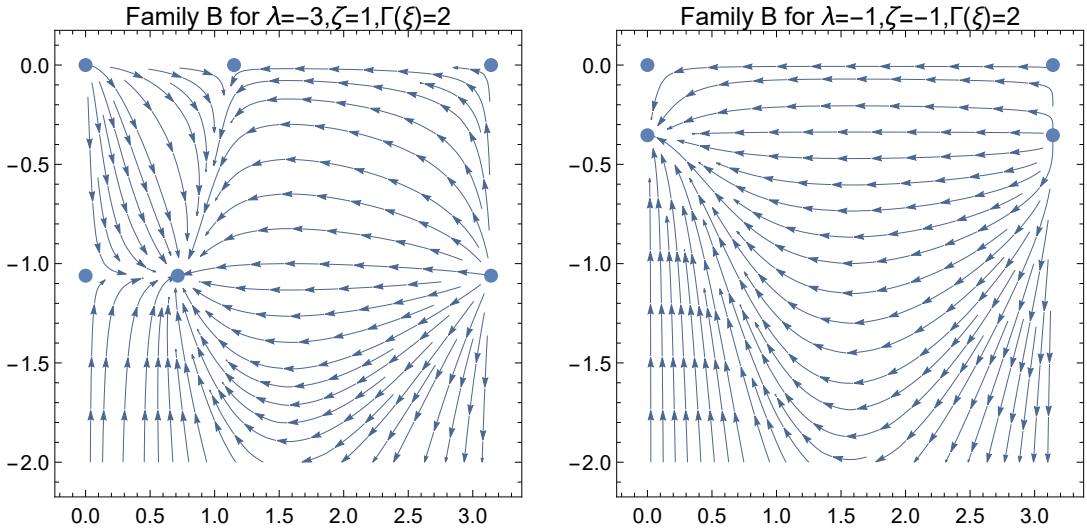} \newline%
\caption{Phase-space diagrams in the two-dimensional space $\left\{
\omega,\xi\right\}  $ for the dynamical system of Family B. Left plot is for
$\left\{  \lambda,\zeta,\Gamma^{\left(  \xi\right)  }\right\}  =\left(
-3,1,2\right)  $ while Right plot is for $\left(  -1,-1,2\right)  $. The
points in the plots are the critical points in the specific region of the
variables. }%
\label{fambb}%
\end{figure}

In Fig. \ref{fambb} we present the phase-space diagram of the two-dimensional
system in the variables $\left\{  \omega,\xi\right\}  $ for different values
of the free parameters and for $\Gamma^{\left(  \xi\right)  }\left(
\xi\right)  $ be a constant, the latter means $Y\left(  \phi\right)  =Y_{0}%
\ln\left(  Y_{1}-Y_{0}e^{-\frac{\lambda\phi}{2}}\right)  $.

\subsection{Family C}

The third system of our consideration is consisted by the differential
equations (\ref{ds.01}), (\ref{ds.02}) and (\ref{ds.05}). The latter dynamical
system admits the following stationary points%
\begin{align}
C_{1}^{\pm}  &  =\left(  \pm1,0,\zeta_{0}\right)  ,~\Gamma^{\left(
\zeta\right)  }\left(  \zeta_{0}\right)  =0,\\
C_{2}^{\pm}  &  =\left(  -\frac{2\sqrt{2}\lambda\pm\sqrt{\xi^{2}\left(
3\left(  4+\xi^{2}\right)  -2\lambda^{2}\right)  }}{\sqrt{3}\left(  4+\xi
^{2}\right)  },1-\left(  C_{2\left(  x\right)  }^{\pm}\right)  ^{2},\zeta
_{0}\right)  ,~~\Gamma^{\left(  \zeta\right)  }\left(  \zeta_{0}\right)  =0,\\
C_{3}^{\pm}  &  =\left(  0,\pm1,\frac{\sqrt{6}\xi-2\lambda}{4}\right)  ,
\end{align}
defined in the space $C=\left(  C_{x},C_{y},C_{\zeta}\right)  $.

The physical properties of the solutions at points $C_{1}^{\pm},~C_{2}^{\pm}$
and $C_{3}^{\pm}$ are described by that of points $A_{1}^{\pm},~A_{2}^{\pm}$
and $~B_{4}^{\pm}$ respectively, where $C_{2}^{\pm}$ should seen as the
special case of $~A_{2}^{\pm}$ with $\zeta=0$. That is, points $C_{1}^{\pm
},~C_{2}^{\pm}$ describe scaling solutions while points $C_{3}^{\pm}$ describe
de Sitter universes.

\subsubsection{Stability analysis}

In order to study the stability of the stationary points we prefer to work on
the variables $\left\{  \omega,\zeta\right\}  $.

The eigenvalues of points $C_{1}^{\pm}$ are calculated
\begin{equation}
e_{1}\left(  C_{1}^{\pm}\right)  =1\pm\left(  \sqrt{\frac{2}{3}}\zeta
_{0}+\frac{\lambda}{\sqrt{6}}\right)  ,~e_{2}\left(  C_{1}^{\pm}\right)
=\pm\sqrt{\frac{2}{3}}\Gamma_{,\zeta}^{\left(  \zeta\right)  }\left(
0\right)  ,
\end{equation}
from where we infer that point $C_{1}^{+}$ is an attractor when $\left\{
\lambda<-\sqrt{6},\Gamma_{,\zeta}^{\left(  \zeta\right)  }\left(  \zeta
_{0}\right)  <0\right\}  $, while $C_{2}^{-}$ is an attractor when $\left\{
\lambda>\sqrt{6},~\Gamma_{,\zeta}^{\left(  \zeta\right)  }\left(  \zeta
_{0}\right)  >0\right\}  $.

As far as the linearized systems around points $C_{2}^{\pm}$ are concerned the
eigenvalues are found to be
\begin{align}
e_{1}\left(  C_{2}^{\pm}\right)   &  =-\frac{8\lambda\text{$\Gamma$}^{\left(
\zeta\right)  \prime}(\text{$\zeta$}_{0})+8\zeta_{0}\lambda-4\lambda^{2}%
+6\xi^{2}+2\xi Y\left(  \xi,\lambda\right)  \text{$\Gamma$}^{\left(
\zeta\right)  \prime}(\text{$\zeta$}_{0})\pm2\zeta_{0}\xi Y\left(  \xi
,\lambda\right)  \mp\lambda\xi Y\left(  \xi,\lambda\right)  +24+\Delta^{2}%
}{12\left(  \xi^{2}+4\right)  },\\
e_{2}\left(  C_{2}^{\pm}\right)   &  =-\frac{8\lambda\text{$\Gamma$}^{\left(
\zeta\right)  \prime}(\text{$\zeta$}_{0})+8\zeta_{0}\lambda-4\lambda^{2}%
+6\xi^{2}+2\xi Y\left(  \xi,\lambda\right)  \text{$\Gamma$}^{\left(
\zeta\right)  \prime}(\text{$\zeta$}_{0})+2\zeta_{0}\xi Y\left(  \xi
,\lambda\right)  -\lambda\xi Y\left(  \xi,\lambda\right)  +24-\Delta^{2}%
}{12\left(  \xi^{2}+4\right)  },
\end{align}
where
\begin{align}
\Delta^{2}\left(  \zeta_{0},\xi,\lambda\right)   &  =\left(  8\zeta_{0}%
\lambda-4\lambda^{2}+6\xi^{2}+2\text{$\Gamma_{,\zeta}^{\left(  \zeta\right)
}$}(\text{$\zeta$}_{0})(4\lambda\pm\xi Y)\pm2\zeta_{0}\xi Y\mp\lambda\xi
Y+24\right)  ^{2}+\nonumber\\
&  2\lambda^{3}\left(  \xi^{2}-4\right)  -4\lambda^{2}\left(  \zeta_{0}\left(
\xi^{2}-4\right)  \pm\xi Y\left(  \zeta_{0},\xi,\lambda\right)  \right)
+\nonumber\\
&  -16\text{$\Gamma_{,\zeta}^{\left(  \zeta\right)  }$}(\text{$\zeta$}%
_{0})\lambda\left(  -3\xi^{4}\pm8\zeta_{0}\xi Y+48\right)  +3\xi\left(
\xi^{2}+4\right)  (2\zeta_{0}\xi\pm Y),
\end{align}
and $Y\left(  \xi,\lambda\right)  =\sqrt{6\left(  4+\xi^{2}\right)
-4\lambda^{2}}$. The stability conditions for that specific point will be
determined in a specific application latter.

Finally, the eigenvalues of the linearized system at points $C_{3}^{\pm}$ are%
\begin{equation}
e_{1}\left(  C_{3}^{\pm}\right)  =-\frac{1}{6}\left(  3+\sqrt{9-24\Gamma
^{\left(  \zeta\right)  }\left(  \zeta_{0}\right)  }\right)  ,~e_{2}\left(
C_{3}^{\pm}\right)  =-\frac{1}{6}\left(  3-\sqrt{9-24\Gamma^{\left(
\zeta\right)  }\left(  \zeta_{0}\right)  }\right)  ,
\end{equation}
where $\zeta_{0}=\frac{\sqrt{6}\xi\left(  \zeta_{0}\right)  -2\lambda}{4}$.
Hence, points $C_{3}^{\pm}$ are stable when $0<\Gamma^{\left(  \zeta\right)
}\left(  \zeta_{0}\right)  <\frac{3}{8}$.

\begin{figure}[ptb]
\centering\includegraphics[width=0.8\textwidth]{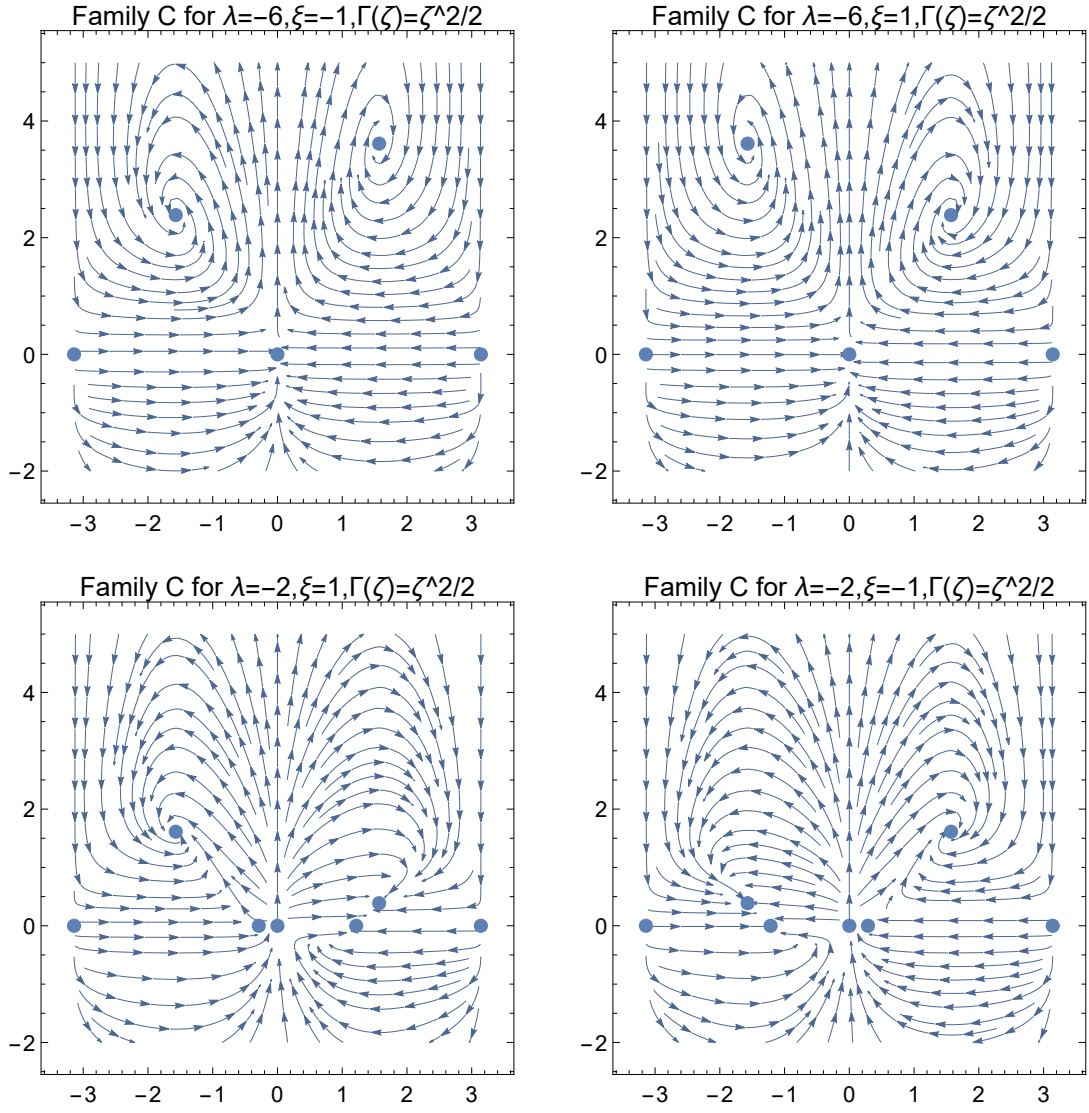} \newline%
\caption{Phase-space diagrams in the two-dimensional space $\left\{
\omega,\zeta\right\}  $ for the dynamical system of Family C. Plots are for
different values of the free-parameters as presented in the labels. }%
\label{famcc}%
\end{figure}

The phase-space diagram of the two-dimensional system in the variables
$\left\{  \omega,\zeta\right\}  $ is presented in Fig. \ref{famcc} for various
values of the free parameters $\left\{  \lambda,\xi\right\}  $ and for
$\Gamma^{\left(  \zeta\right)  }\left(  \zeta\right)  =\Gamma_{0}\zeta^{2}%
$,~that is, $W\left(  \phi\right)  =W_{0}x^{-\Gamma_{0}}$.

\subsection{Family D}

For the fourth system of our consideration, in which $\lambda\neq const$, and
from the system of equations (\ref{ds.01}), (\ref{ds.02}) and (\ref{ds.03}) we
find the stationary points $D=\left(  D_{x},D_{y},D_{\lambda}\right)  $ as
follows%
\begin{align}
D_{1}^{\pm}  &  =\left(  1,0,0\right)  ,\\
D_{2}^{\pm}  &  =\left(  1,0,\lambda_{0}\right)  ~,~\lambda_{0}\left(
1-\Gamma^{\left(  \lambda\right)  }\left(  \lambda_{0}\right)  \right)
=\zeta\left(  \lambda_{0}\right)  ,\\
D_{3}^{\pm}  &  =\left(  \frac{-2\sqrt{2}\left(  2\zeta+\lambda_{0}\right)
\pm\sqrt{3\left(  4+\xi^{2}\right)  \xi^{2}-2\left(  \left(  2\zeta
+\lambda_{0}\right)  \xi\right)  ^{2}}}{\sqrt{3}\left(  4+\xi^{2}\right)
},1-\left(  D_{3\left(  x\right)  }^{\pm}\right)  ^{2},\lambda_{0}\right)
~,~\lambda_{0}\left(  1-\Gamma^{\left(  \lambda\right)  }\left(  \lambda
_{0}\right)  \right)  =\zeta\left(  \lambda_{0}\right) \\
D_{4}^{\pm}  &  =\left(  0,\pm1,\lambda_{0}\right)  ~,~\lambda_{0}=-\left(
2\zeta+\sqrt{\frac{3}{2}}\xi\right)  ~,\\
D_{5}^{\pm}  &  =\left(  \frac{-4\sqrt{2}\zeta\pm\sqrt{3\left(  4+\xi
^{2}\right)  \xi^{2}-8\left(  \zeta\xi\right)  ^{2}}}{\sqrt{3}\left(
4+\xi^{2}\right)  },1-\left(  D_{5\left(  x\right)  }^{\pm}\right)
^{2},0\right)  .
\end{align}

We observe that there are five sets of stationary points with physical
properties as described by points $B_{1}^{\pm},~B_{2}^{\pm},~B_{3}^{\pm
},~B_{4}^{\pm}$ and $B_{5}^{\pm}$ respectively. We proceed by studying the
stability of the stationary points.

\subsubsection{Stability analysis}

As in the previous families of stationary points we study the stability of the
stationary points for the two dimensional system in the variables$~\left\{
\omega,\lambda\right\}  $.

For the points $D_{1}^{\pm}$ the eigenvalues are calculated
\begin{equation}
e_{1}\left(  D_{1}^{\pm}\right)  =\pm\sqrt{\frac{2}{3}}\zeta\left(  0\right)
~,~e_{2}\left(  D_{1}^{\pm}\right)  =1\pm\frac{\sqrt{6}}{3}\zeta\left(
0\right)  ,
\end{equation}
from where we infer that point $D_{1}^{+}$ is an attractor when $\zeta\left(
0\right)  <-\frac{3}{2}$, while $D_{1}^{-}$ is an attractor when $\zeta\left(
0\right)  >\frac{3}{2}$.

The eigenvalues of the linearized system at points $D_{2}^{\pm}$ are
\begin{align*}
e_{1}\left(  D_{2}^{\pm}\right)   &  =1\pm\frac{\sqrt{6}}{6}\left(
2\zeta\left(  \lambda_{0}\right)  +\lambda_{0}\right)  ~,~\\
e_{2}\left(  D_{2}^{\pm}\right)   &  =-\sqrt{\frac{2}{3}}\left(  \zeta\left(
\lambda_{0}\right)  \mp\lambda_{0}\left(  \lambda_{0}\Gamma_{,\lambda
}^{\left(  \lambda\right)  }\left(  \lambda_{0}\right)  +\zeta_{\lambda
}\left(  \lambda_{0}\right)  \right)  \right)  .
\end{align*}

Hence, point $D_{2}^{+}$ is an attractor when \ $\left(  2\zeta\left(
\lambda_{0}\right)  +\lambda_{0}\right)  <-\sqrt{6}$ and $\zeta\left(
\lambda_{0}\right)  >\lambda_{0}\left(  \lambda_{0}\Gamma_{,\lambda}^{\left(
\lambda\right)  }\left(  \lambda_{0}\right)  +\zeta_{\lambda}\left(
\lambda_{0}\right)  \right)  $, while point $D_{2}^{-}$ is an attractor when
\ $\left(  2\zeta\left(  \lambda_{0}\right)  +\lambda_{0}\right)  >\sqrt{6}$
and $\zeta\left(  \lambda_{0}\right)  >-\lambda_{0}\left(  \lambda_{0}%
\Gamma_{,\lambda}^{\left(  \lambda\right)  }\left(  \lambda_{0}\right)
+\zeta_{\lambda}\left(  \lambda_{0}\right)  \right)  $.

As far as the points $D_{3}^{\pm}$ are concerned, the eigenvalues are%
\begin{align}
e_{1}\left(  D_{3}^{\pm}\right)   &  =\frac{-2\left(  3\xi^{2}-2\left(
\lambda_{0}^{2}-6\right)  +8\zeta\left(  \lambda_{0}+\zeta\right)  \right)
\mp\left(  \lambda_{0}+2\zeta\right)  \sqrt{2\left(  3\xi^{2}-2\left(
\lambda_{0}^{2}-6\right)  +8\zeta\left(  \lambda_{0}+\zeta\right)  \right)  }%
}{6\left(  4+\xi^{2}\right)  },\\
e_{2}\left(  D_{3}^{\pm}\right)   &  =-\frac{2\left(  2\left(  \lambda
_{0}+2\zeta\right)  ^{2}-3\xi^{2}\right)  \left(  \lambda_{0}\left(
\lambda_{0}\Gamma_{,\lambda}^{\left(  \lambda\right)  }\left(  \lambda
_{0}\right)  +\zeta_{\lambda}\left(  \lambda_{0}\right)  \right)
-\zeta\right)  }{3\left(  4\lambda_{0}+8\zeta\pm\xi\sqrt{2\left(  3\xi
^{2}-2\left(  \lambda_{0}^{2}-6\right)  +8\zeta\left(  \lambda_{0}%
+\zeta\right)  \right)  }\right)  },
\end{align}
in which $\xi=\xi\left(  \lambda_{0}\right)  $ and $\zeta=\zeta\left(
\lambda_{0}\right)  $. \ 

\begin{figure}[ptb]
\centering\includegraphics[width=0.8\textwidth]{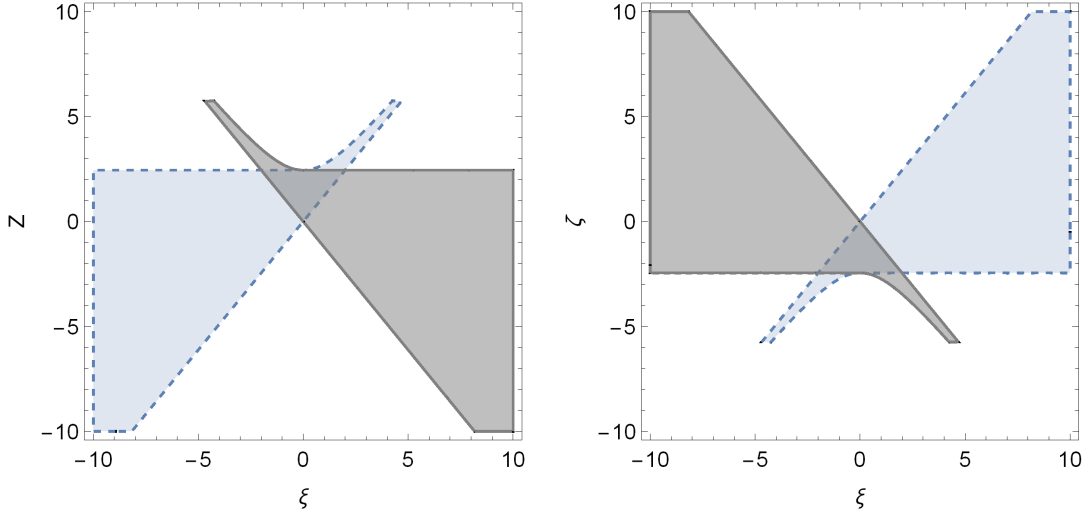} \newline%
\caption{Region plot in the space of variables $\left\{  \xi,Z\right\}  $
where the exact solutions at points $D_{3}^{\pm}$ are stable. Blue area
corredpond to the values where point $D_{3}^{+}$ is stable, while gray area is
for point $D_{3}^{-}$. Left Fig. is for $\Gamma_{\lambda}^{\left(
\lambda\right)  }\left(  \lambda_{0}\right)  >0$ while right Fig. is for
$\Gamma_{\lambda}^{\left(  \lambda\right)  }\left(  \lambda_{0}\right)  <0$. }%
\label{figd3}%
\end{figure}

In order to simplify the stability conditions, we need to specify the unknown
functions $\xi\left(  \lambda\right)  ,~\zeta\left(  \lambda\right)  $ and
$\Gamma^{\left(  \lambda\right)  }\left(  \lambda\right)  $. In the specific
case where $\xi,~\zeta$ are constants, it follows that $D_{3}^{+}$ is an
attractor when $\Gamma_{\lambda}^{\left(  \lambda\right)  }\left(  \lambda
_{0}\right)  >0:\left\{  \lambda_{0}^{2}\Gamma_{\lambda}^{\left(
\lambda\right)  }\left(  \lambda_{0}\right)  >\zeta,Z<\sqrt{6},\xi
<-2,\left\vert \xi\right\vert <\frac{2Z}{\sqrt{6}}\right\}  $ or $\left\{
0<\xi<\frac{2Z}{\sqrt{6}},\sqrt{6\left(  4+\xi^{2}\right)  }>2Z\right\}  $ or
$\left\{  -2<\xi<0,\xi<-\frac{2Z}{\sqrt{6}}\right\}  $; $\Gamma_{\lambda
}^{\left(  \lambda\right)  }\left(  \lambda_{0}\right)  <0:$ $\left\{
\frac{2Z}{\sqrt{6}}<\xi<0,\sqrt{6\left(  4+\xi^{2}\right)  }>-2Z\right\}  $ or
$\left\{  -\frac{2Z}{\sqrt{6}}<\xi<2\right\}  $; $\left\{  Z>-\sqrt{6}%
,0<\xi<2,\xi<-\frac{2Z}{\sqrt{6}}\right\}  $; $\left\{  \xi_{0}>2,\xi
_{0}>\frac{2Z}{\sqrt{6}}\right\}  $ where $Z=2\zeta+\lambda_{0}$. In addition,
point $D_{3}^{-}$ is an attractor when $\Gamma_{\lambda}^{\left(
\lambda\right)  }\left(  \lambda_{0}\right)  >0:\left\{  Z<\sqrt{6}%
,0<\xi<\frac{2Z}{\sqrt{6}}\right\}  $or $\left\{  Z<\sqrt{6},2<\xi<-\frac
{2Z}{\sqrt{6}}\right\}  $or $\left\{  \xi<-\frac{2Z}{\sqrt{6}},2Z<\sqrt
{6\left(  4+\xi^{2}\right)  }<0\right\}  $ or $\left\{  \frac{2Z}{\sqrt{6}%
}<\xi<2,\right\}  ;$ $\Gamma_{\lambda}^{\left(  \lambda\right)  }\left(
\lambda_{0}\right)  <0:\left\{  \xi_{0}<-\frac{2Z}{\sqrt{6}},\right\}  $ or
$\left\{  -2<\xi_{0}<\frac{2Z}{\sqrt{6}}\right\}  $ or $\left\{  Z>-\sqrt
{6},\xi<-2\right\}  $ or $\left\{  \xi>0,\sqrt{6\left(  4+\xi^{2}\right)
}>-2Z\right\}  $ or~$\left\{  Z>-\sqrt{6},\frac{2Z}{\sqrt{6}}<\xi<0\right\}
.$ The latter regions plots are presented in Fig. \ref{figd3}.

For the points $D_{4}^{\pm}$ we find%
\begin{equation}
e_{1}\left(  D_{4}^{\pm}\right)  =\frac{1}{6}\left(  -3-\sqrt{3}\sqrt
{\Delta\left(  D_{4}^{\pm}\right)  }\right)  ~,~e_{2}\left(  D_{4}^{\pm
}\right)  =\frac{1}{6}\left(  -3+\sqrt{3}\sqrt{\Delta\left(  D_{4}^{\pm
}\right)  }\right)  ,~
\end{equation}
with
\begin{align}
\Delta\left(  D_{4}^{\pm}\right)   &  =3+4\lambda_{0}^{2}-4\lambda_{0}\left(
\lambda_{0}\Gamma^{\left(  \lambda\right)  }\left(  \lambda_{0}\right)
+\zeta\left(  \lambda_{0}\right)  \right)  +\nonumber\\
&  -8\lambda\zeta_{\lambda}\left(  \lambda_{0}\right)  \left(  \lambda
_{0}\left(  \Gamma^{\left(  \lambda\right)  }\left(  \lambda_{0}\right)
-1\right)  +\zeta\left(  \lambda_{0}\right)  \right)  +\nonumber\\
&  \mp2\sqrt{6}\lambda\left(  \lambda\left(  \Gamma^{\left(  \lambda\right)
}\left(  \lambda_{0}\right)  -1\right)  +\zeta\left(  \lambda_{0}\right)
\right)  \xi_{\lambda}\left(  \lambda_{0}\right)  ,
\end{align}
we can not extract additional conditions for the stability of points
$\Delta\left(  D_{4}^{\pm}\right)  $ without considering special forms of the
unknown functions.

\begin{figure}[ptb]
\centering\includegraphics[width=0.4\textwidth]{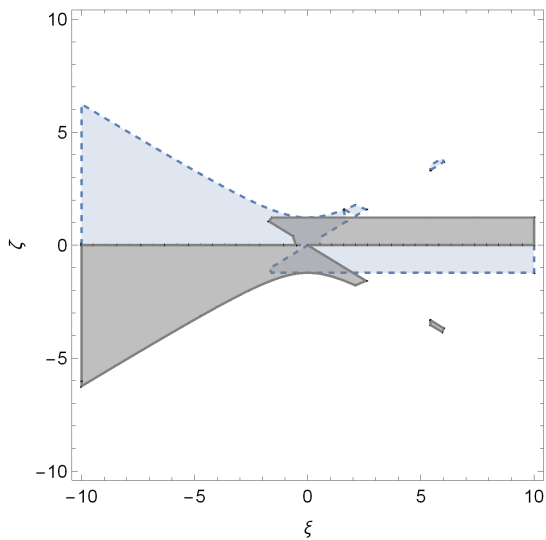} \newline%
\caption{Region plot in the space of variables $\left\{  \xi,\zeta\right\}  $
where the exact solutions at points $D_{5}^{\pm}$ are stable. Blue area
corredpond to the values where point $D_{5}^{+}$ is stable, while gray area is
for point $D_{5}^{-}$. }%
\label{figd5}%
\end{figure}

The eigenvalues at points $D_{5}^{\pm}$ are%
\begin{align}
e_{1}\left(  D_{5}^{\pm}\right)   &  =-\frac{8\zeta\mp\sqrt{2}\sqrt{3\left(
4+\xi^{2}\right)  \xi^{2}-8\left(  \zeta\xi\right)  ^{2}}}{3\left(  4+\xi
^{2}\right)  }\zeta,\\
e_{2}\left(  D_{5}^{\pm}\right)   &  =\frac{\left(  3\left(  4+\xi^{2}\right)
-8\left(  \zeta\right)  ^{2}\right)  \pm\sqrt{2}\zeta\sqrt{3\left(  4+\xi
^{2}\right)  \xi^{2}-8\left(  \zeta\xi\right)  ^{2}}}{3\left(  4+\xi
^{2}\right)  }.
\end{align}
Therefore, point $D_{5}^{+}$ is an attractor when~$\left\{  \xi=0,0<\zeta
<\frac{\sqrt{6}}{2}\right\}  $;$~\zeta>-\frac{\sqrt{6}}{2}:\left\{
-2<\xi<0,\zeta<\frac{\sqrt{6}}{4}\xi\right\}  $ or~$\left\{  \xi
>0,\zeta<0\right\}  $; $\zeta<\frac{\sqrt{6\left(  4+\xi^{2}\right)  }}%
{4}:\left\{  \zeta>0,\xi<0\right\}  $ or $\left\{  0<\xi<\frac{4}{\sqrt{6}%
}\zeta\right\}  $. On the other hand, point $D_{5}^{-}$ is an attractor when
~$\xi>0:\left\{  0<\zeta<\frac{3}{2}\right\}  $;$~\left\{  4\zeta+\sqrt
{6}\left(  4+\xi^{2}\right)  >0,4\zeta+\sqrt{6}\xi<0\right\}  $ or
$\xi<0:\left\{  4\zeta+\sqrt{6\left(  4+\xi^{2}\right)  }>0,~\zeta<0\right\}
$ or $\left\{  4\zeta+\sqrt{6}\xi>0,~2\zeta<\sqrt{6}\right\}  \,.$ Recall that
in the latter, $\xi$ and $\zeta$ correspond to $\xi\left(  0\right)  $ and
$\zeta\left(  0\right)  $. In Fig. \ref{figd5} we present the regions in the
space $\left\{  \xi\left(  0\right)  ,\zeta\left(  0\right)  \right\}  $ where
points $D_{5}^{\pm}$ are attractors.

\begin{figure}[ptb]
\centering\includegraphics[width=0.8\textwidth]{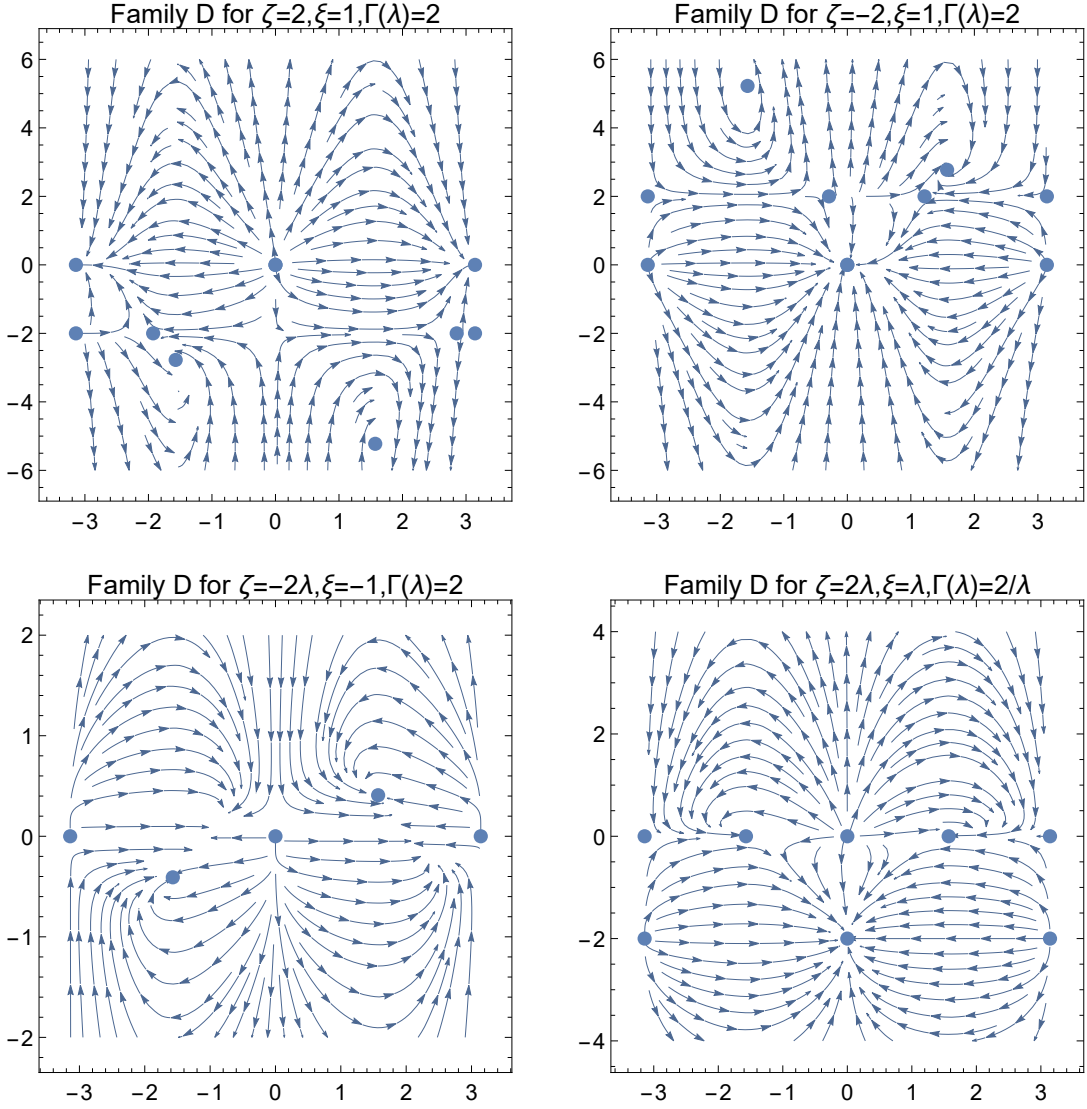} \newline%
\caption{Phase-space diagrams in the two-dimensional space $\left\{
\omega,\lambda\right\}  $ for the dynamical system of Family D. Plots are for
different forms of the free functions $\left\{  \zeta\left(  \lambda\right)
,\xi\left(  \lambda\right)  ,\Gamma^{\left(  \lambda\right)  }\left(
\lambda\right)  \right\}  $. }%
\label{famdd}%
\end{figure}

The phase-space diagram of the two-dimensional system in the variables
$\left\{  \omega,\lambda\right\}  $ is presented in Fig. \ref{famdd} for
various functional forms of the free functions $\left\{  \zeta\left(
\lambda\right)  ,\xi\left(  \lambda\right)  ,\Gamma^{\left(  \lambda\right)
}\left(  \lambda\right)  \right\}  $.

\section{Conclusions}

\label{sec5}

We preformed an extended analysis on the dynamics of the Einstein-Aether
cosmology with a scalar field coupled to the aether field by generalized the
analysis presented in \cite{anaether}. Such an analysis is important in order
to understand the viability of the Einstein-\ae ther scalar field cosmology,
as also to understand the contribution of new interaction terms, between the
scalar field and the aether field, in the gravitational field equations. In
order to study the dynamics the cosmological evolution we studied the field
equations in dimensionless form by using the $\theta-$normalization, and we
determined the stationary points. Each stationary point describe a specific
phase in the cosmological history of the model.

We assumed that the scalar field and the aether field contribute in the
gravitational integral a potential term of the form $V\left(  \theta
,\phi\right)  =U\left(  \phi\right)  +Y\left(  \phi\right)  \theta+\frac{1}%
{3}\left(  W^{2}\left(  \phi\right)  -1\right)  \theta^{2},~$were $U\left(
\phi\right)  ,~Y\left(  \phi\right)  $ and $W\left(  \phi\right)  $ are
arbitrary function. When $W^{2}\left(  \phi\right)  =1,$ or in general when
$W\left(  \phi\right)  =const$, the analysis of \cite{anaether} are recovered.
In addition when $W\left(  \phi\right)  =const$, $U\left(  \phi\right)
=V_{0}e^{-\lambda\phi}$ and $Y\left(  \phi\right)  =Y_{0}e^{-\frac{1}%
{2}\lambda\phi}$ the results of \cite{ra1} are recovered. Indeed when the
function $W\left(  \phi\right)  =const$, then in \cite{anaether} it was found
that that there are three families of stationary points, while in our
consideration for the arbitrary function $W\left(  \phi\right)  $ there are
four families of stationary points.

By writing the field equations with the use of the Einstein tensor, we observe
that the contribution of $W\left(  \phi\right)  $ is similar with the coupling
function of the scalar field with gravity in Scalar-tensor theories. While in
our model the scalar field is only coupled with the aether field, however
there is an undirected coupling with the gravity. In particular $W\left(
\phi\right)  $ can be used to define an effective varying gravitational
\textquotedblleft constant\textquotedblright\ $k_{eff}=W^{-2}\left(
\phi\right)  $.

The first family of stationary points in \cite{anaether}, namely Family \={A},
it is consisted by two sets of stationary points which describe scaling
solutions. The stationary points of Family \={B} are four pairs of stationary
points, while the third family of stationary points, namely Family \={C}, are
again four pairs of stationary points. \ It is important to mention that in
\cite{anaether} it was assumed that parameter $y$ is always positive.

In the model of this work, the models of Families A,~B and D can be seen as
the generalized Families of \={A}, \={B} and \={D} respectively. On the other
hand, Family C describes new stationary points provided by our model and
specifically by the nonconstant function $W\left(  \phi\right)  $.

Family A consists two pair of stationary points which describe scaling
solutions as the points of Family \={A}. The stationary points of Families
B~and C consist five pairs of points, where the four pairs describe scaling
solutions and only one pair describes de Sitter universe. However all the
points have their equivalent in Families \={B} and \={D} by using the
presentation of \cite{anaether}. Because the dimension of the system is
different from the case where $W\left(  \phi\right)  =const.$ the stability
conditions are modified as also the physical variables, however when $W\left(
\phi\right)  =const,~$we end up with the same results of \cite{anaether}.
Family C for the model of our consideration admits six stationary points in
three pairs. The two pairs describe scaling universes while the third pair of
points describe de Sitter universes. Recall that the de Sitter solution it is
supported by the cosmological observations to be the attractor of the late
time cosmic acceleration phase of the universe.

From our analysis we found that the introduction of the new potential term in
the field equations modifies the dynamics. However, while someone will expect
the stationary points to be different we found that there is an one to one
correspondence between all the stationary points for the $W\left(
\phi\right)  =const$ and\ the case where $W\left(  \phi\right)  $ it is an
arbitrary function. The only new stationary points\ are those of Family C.

Consequently, when $V\left(  \theta,\phi\right)  =U\left(  \phi\right)
+Y\left(  \phi\right)  \theta$ or $V\left(  \theta,\phi\right)  =U\left(
\phi\right)  +Y\left(  \phi\right)  \theta+\bar{W}^{2}\left(  \phi\right)
\theta^{2}$, the cosmological history has a similar evolution. By the results
of this work we can conclude that the model $V\left(  \theta,\phi\right)
=U\left(  \phi\right)  +Y\left(  \phi\right)  \theta+\bar{W}^{2}\left(
\phi\right)  \theta^{2}$ it can describe the basic cosmological history, a
similar result which is expected and for the general model $V\left(
\theta,\phi\right)  $, since more degrees of freedom are introduced. Of course
the latter conclusion it follows from the evolution of the solution
trajectories of the field equations. Recall that when $Y\left(  \phi\right)
=0$, that is, $V\left(  \theta,\phi\right)  =U\left(  \phi\right)  +\bar
{W}^{2}\left(  \phi\right)  \theta^{2}$ our model describes also the one
considered by Kanno et al \cite{soda1}.

The field equations of the model with $Y\left(  \phi\right)  =0$ in the
dimensionless variables are these with $\xi=0$. Therefore, only the stationary
points of families $A,$ $C$ and $D$ exist with the additional constraint~$\xi
=0$. Consequently, we can conclude that the introduction of the function
$Y\left(  \phi\right)  $ enriches the evolution of the cosmological history.

Let us now discuss the physical interpretation of the critical points.
\ Points $A_{1}^{\pm},~A_{2}^{\pm}$ describe scaling solutions in general,
however for specific values of the free parameters these solutions can
describe also de Sitter spacetimes. Consequently, for specific ranges of the
free parameters the points of family $A$ can describe an unstable scaling
solution which describes the inflationary era, as also a future de Sitter
attractor. The situation is similar and for the rest families of critical
points. Families B, C~and D\ can admit more than two sets of critical points,
but that does not mean that all those solutions can play role in the
cosmological evolution, since the cosmological evolution described by the
field equations it depends on the initial conditions, as it is demonstrated by
the phase space diagrams presented in Figs. \ref{fambb}, \ref{famcc} and
\ref{famdd}. Recall that Einstein-aether have been tested as a dark energy
alternative in \cite{in2}.

Nevertheless, if someone would like to describe the complete cosmological
history then radiation and dust fluids should be introduced in the field
equations such that to describe the radiation and the matter dominated epochs.
By performing a similar analysis in a scenario with more matter sources in the
cosmological model, we except to find critical points where the radiation
fluid or the dust fluid contribute or dominate in the cosmological evolution,
such that to describe the radiation and matter eras. Further, the existence of
new critical points where all the fluid sources contribute are expected to
exist, similarly with the quintessence and the scalar tensor theories, for
more details we refer the reader in Appendix A. On the other hand, we can
require the scaling solutions that we found before to describe the additional
eras of the cosmological evolution, as for example the Brans-Dicke provides an
ideal gas solution or the $f\left(  R\right)  $-theory which provides a
radiation epoch \cite{cop2,ffr,ffr2}.

Supplementary, we remark that there exist and other exact solutions for the
field equations (\ref{l0.1}), (\ref{l0.2}) and (\ref{l0.3}) except from the
scaling and de Sitter solutions. As it was discussed in \cite{anaether}
because the unknown functions are more than the equation of motions someone
can construct various analytical solutions which can describe well studied
cosmological solutions. For instance, if we assume $W\left(  \phi\right)
=W_{0}\phi\left(  t\right)  ,~\phi\left(  t\right)  =\phi_{0}t$ and
$\theta\left(  t\right)  =\theta_{1}\coth\left(  \theta_{0}t\right)  $~such
that to describe the $\Lambda$-cosmology, it follows necessarily $U\left(
\phi\right)  =$ $\frac{1}{6}\left(  2W_{0}^{2}\phi^{2}\left(  \theta
_{1}\right)  ^{2}\coth^{2}\left(  \theta_{0}\phi\right)  -3\right)  $ and
$Y\left(  \phi\right)  =-\frac{2}{3}W_{0}^{2}\phi\theta_{1}\coth\left(
\theta_{0}\phi\right)  \,$, however for different functional forms of
$\phi\left(  t\right)  $, the $\Lambda$-cosmology can be recovered but for
different functions $U\left(  \phi\right)  ,~Y\left(  \phi\right)  $. The main
difference between the various classical solutions which can be found is,
which is the attractor of the solution, the scaling solution $\theta\left(
t\right)  \simeq t^{-1}$ or the de Sitter solution $\theta\left(  t\right)
\simeq\theta_{1}$.

Additional analysis which should include cosmological observations as also to
study the effects of the interaction term in the perturbation level should be
performed. However, such analysis is beyond the purpose of this work.

From the above results we see that maybe it is not necessary to introduce more
nonlinear interactions between the scalar field and the aether field, at least
in the context of the cosmological solutions.

\begin{acknowledgments}
A.P. thanks Dr. Alex Giacomini and the Universidad Austral de Chile for the
hospitality provided while this work was performed.
\end{acknowledgments}

\appendix

\section{Stationary points in the presence of matter}

In this Appendix, we consider the existence of a dust fluid in the
cosmological model, which is not interacting with the aether or the scalar
fields.\ The only field equation which is modified it is equation\ (\ref{l0.1}%
) and becomes
\begin{equation}
\frac{1}{3}W^{2}\left(  \phi\right)  \theta^{2}=\frac{1}{2}\dot{\phi}%
^{2}+U\left(  \phi\right)  +\rho_{m}%
\end{equation}
where $\rho_{m}$ is the energy density of the fluid.

Because the second-order differential equations (\ref{l0.2}), (\ref{l0.3})
remain the same, when we introduce the pressureless fluid, we infer that the
field equations in the dimensionless variables (\ref{ds.01})-(\ref{ds.05}) are
the same. However, the constraint equation (\ref{ds.07}) reads
\begin{equation}
\Omega_{m}=1-x^{2}-y^{2}%
\end{equation}
where now the new variable $\Omega_{m}=\frac{3\rho_{m}}{W^{2}\left(
\phi\right)  \theta^{2}}$, describes the energy density of the dust fluid
source~constrained in the range $0\leq\Omega_{m}\leq1.$

There are two main differences with the results presented in Section
\ref{sec4}. Firstly, the critical points are not necessary points on the
unitary circle in the two dimensional space $\left\{  x,y\right\}  $, but they
are located in the unitary disk with center the point $\left(  0,0\right)  $.
Moreover, because of the introduction of the extra variable $\Omega_{m},$ the
dimension of the dynamical system has been increased by one. We present the
additional points for the field equations where $\Omega_{m}$ is different from
zero. However, we do not perform a complete analysis, by mean, we are not
calculate the new stability conditions.

In Family A, the additional critical points are4%
\[
A_{1}^{m}=\left(  -2\sqrt{\frac{2}{3}}\zeta,0\right)  ~,~A_{2\left(
\pm\right)  }^{m}=\left(  -\frac{1}{\lambda}\sqrt{\frac{3}{2}},\frac{-\sqrt
{3}\xi\pm\sqrt{3\left(  4+\xi^{2}\right)  -16\zeta\lambda}}{2\sqrt{2}\lambda
}\right)
\]
from where the energy density of the dust fluid is calculated%
\begin{equation}
\Omega_{m}\left(  A_{1}^{m}\right)  =1-\frac{8}{3}\zeta^{2},
\end{equation}
and%
\begin{equation}
\Omega_{m}\left(  A_{2\left(  \pm\right)  }^{m}\right)  =1-\frac{3}%
{\lambda^{2}}+\frac{2\zeta}{\lambda}-\frac{3\xi^{2}}{4\lambda^{2}}\pm
\frac{\sqrt{3}\xi}{4\lambda^{2}}\sqrt{3\left(  4+\xi^{2}\right)
-16\zeta\lambda}.
\end{equation}

Point $A_{1}^{m}~$is physical accepted when $\left\vert \zeta\right\vert
\leq\frac{1}{2}\sqrt{\frac{3}{2}}$, while for points $~A_{2\left(  \pm\right)
}^{m}$ when~$3\left(  4+\xi^{2}\right)  -16\zeta\lambda\geq0$ and $0\leq
\Omega_{m}\left(  A_{2\left(  \pm\right)  }^{m}\right)  \leq1$. Hence, for
point~$A_{2\left(  +\right)  }^{m}$ we find the constraints $\xi\leq0~$:
$\left\{  \xi=\frac{3\left(  4+\xi^{2}\right)  }{16\lambda},\pm4\lambda
+\sqrt{6\left(  4+\xi^{2}\right)  }=0\right\}  $ or $\left\{  \frac{3\left(
4+\xi^{2}\right)  }{16\lambda}\leq\zeta\leq\frac{6-2\lambda^{2}-\left\vert
\xi\right\vert \sqrt{6-\frac{9}{\lambda^{2}}}\lambda}{4\lambda},4\lambda
+\sqrt{6\left(  4+\xi^{2}\right)  }<0\right\}  $ or $\left\{  \frac
{6-2\lambda^{2}+\left\vert \xi\right\vert \sqrt{6-\frac{9}{\lambda^{2}}%
}\lambda}{4\lambda}\leq\zeta\leq\frac{3\left(  4+\xi^{2}\right)  }{16\lambda
},-4\lambda+\sqrt{6\left(  4+\xi^{2}\right)  }<0\right\}  $, while when
$\xi>0$ :~$\left\{  4\zeta\pm\sqrt{6}=0,\sqrt{6}+2\lambda=0\right\}  $ or
$\left\{  \zeta>\frac{3\left(  4+\xi^{2}\right)  }{16\lambda},2\left(
2\zeta+\lambda\right)  \leq\frac{6+\sqrt{6-\frac{9}{\lambda^{2}}}\lambda\xi
}{\lambda},4\lambda+\sqrt{6\left(  4+\xi^{2}\right)  }\leq0\right\}  $
$\ $or~$\left\{  \zeta\leq\frac{3\left(  4+\xi^{2}\right)  }{16\lambda
},-4\lambda+\sqrt{6\left(  4+\xi^{2}\right)  }\leq0,4\zeta+2\lambda
+\sqrt{6-\frac{9}{\lambda^{2}}}\geq\frac{6}{\lambda^{2}}\right\}  $ or
$\left\{  2\left(  2\zeta+\lambda\right)  \leq\frac{6+\sqrt{6-\frac{9}%
{\lambda^{2}}}\lambda\xi}{\lambda},4\zeta+2\lambda+\sqrt{6-\frac{9}%
{\lambda^{2}}}\geq\frac{6}{\lambda^{2}}\right\}  $~with $\left\{  \lambda
<\pm\frac{\sqrt{6}}{2},\mp4\lambda+\sqrt{6\left(  4+\xi^{2}\right)
}>0\right\}  $.

As far as concerns the physical properties of the exact solutions at those new
critical points, we observe that the dust fluid as also the effective fluid of
the scalar field with the aether field contribute in the cosmological
solution. However, in general for this points we find that the parameter for
the equation of state for the effective fluid of the scalar and the aether
fields is different from zero at these points, which indicates that they are
not tracking solutions, but the exact solution has a two ideal gas contribution.

In Family B we find the critical points%
\begin{align*}
B_{1}^{m}  &  =\left(  A_{1}^{m},0\right)  ~,~B_{2\left(  \pm\right)  }%
^{m}=\left(  A_{2\left(  \pm\right)  }^{m},\xi_{0}\right)  ~,~\\
B_{3}^{m}  &  =\left(  A_{1}^{m},\xi_{0}\right)  ~,~B_{4\left(  \pm\right)
}^{m}=\left(  A_{2\left(  \pm\right)  }^{m},0\right)  ~,
\end{align*}
where $\sqrt{2}\Gamma_{2}\left(  \xi_{0}\right)  \xi_{0}=\lambda$. Easily we
observe that sets of points $\left\{  B_{1}^{m},B_{3}^{m}\right\}  $ and
$\left\{  B_{2\left(  \pm\right)  }^{m},B_{4\left(  \pm\right)  }^{m}\right\}
~$has similar physical properties with points $A_{1}^{m}$ and $A_{2\left(
\pm\right)  }^{m}$ respectively.

For Family C the additional stationary points are%
\[
C_{1}^{m}=\left(  1,0,0\right)  \text{ and }C_{2\left(  \pm\right)  }%
^{m}=\left(  A_{2\left(  \pm\right)  }^{m},0\right)  ,
\]
where $C_{1}^{m}$ describes a universe dominated by the pressureless fluid,
while $C_{2\left(  \pm\right)  }^{m}$ have the same physical properties with
points $A_{2\left(  \pm\right)  }^{m}$. However, for points $C_{2\left(
\pm\right)  }^{m}~$because $\zeta=0$~we find that the exact solutions at
$C_{2\left(  \pm\right)  }^{m}\,\ $describe tracking solutions, that is, the
effective fluid of the scalar and the aether field behaves like the dust fluid.

Finally for Family D the new critical points are found to be%
\begin{align*}
D_{1}^{m}  &  =\left(  A_{1}^{m},0\right)  ~,~D_{2\left(  \pm\right)  }%
^{m}=\left(  A_{2\left(  \pm\right)  }^{m},\xi_{0}\right)  ~,~\\
D_{3}^{m}  &  =\left(  A_{1}^{m},\xi_{0}\right)  ~,~D_{4\left(  \pm\right)
}^{m}=\left(  A_{2\left(  \pm\right)  }^{m},0\right)  ~,
\end{align*}
with one to one physical correspondence with points $B_{1}^{m}-B_{4\left(
\pm\right)  }^{m}$.

In the generic scenario that the additinal matter source has a pressure term
of the form $p_{m}=\left(  \gamma-1\right)  \rho_{m}$ where the limit
$\gamma=1$ correspond to the dust fluid source, the field equations
(\ref{ds.01})-(\ref{ds.05}) in the dimensionless variables are modified as follows%

\begin{align}
\frac{dx}{d\tau}  &  =\frac{1}{6}\left(  x^{2}-1\right)  \left(  3x+2\sqrt
{6}\zeta\right)  -\frac{1}{6}y^{2}\left(  3x+\sqrt{6}\lambda\right)  +\frac
{1}{2}\left(  x^{2}-1\right)  y\xi+\frac{\left(  \gamma-1\right)  }{2}%
x\Omega_{m},\\
\frac{dy}{d\tau}  &  =y^{2}\left(  \left(  1-y^{2}\right)  +\frac{1}%
{3}x\left(  3\left(  x+y\xi\right)  +\sqrt{6}\right)  \left(  \lambda+\sqrt
{6}\zeta\right)  \right)  +y^{2}\left(  \gamma-1\right)  \Omega_{m}\\
\frac{d\lambda}{d\tau}  &  =\sqrt{\frac{2}{3}}x\lambda\left(  \zeta
+\lambda\left(  \Gamma^{\left(  \lambda\right)  }\left(  \lambda\right)
-1\right)  \right)  ,\\
\frac{d\xi}{d\tau}  &  =\frac{\sqrt{3}}{6}x\xi\left(  2\xi\Gamma^{\left(
\xi\right)  }\left(  \xi\right)  -\sqrt{2}\lambda\right)  ,\\
\frac{d\zeta}{d\tau}  &  =\frac{\sqrt{6}}{3}x\Gamma^{\left(  \zeta\right)
}\left(  \zeta\right)  ,
\end{align}
from where we find the same families of stationary points which now depend on
the equation of state parameter for the ideal gas $\gamma$.

\end{document}